\documentclass[
prx,
twocolumn,
amsmath,
amssymb,
floatfix,
footinbib,
bibnotes,
longbibliography
]{revtex4-1}
\pdfoutput=1

\usepackage[colorlinks=true,citecolor=blue,linkcolor=magenta]{hyperref}
\usepackage{amsmath}
\usepackage{graphicx}
\usepackage{changepage}
\usepackage{fancyhdr}
\usepackage{amsthm, amssymb}
\usepackage{floatflt}
\usepackage{float}
\usepackage{array}
\usepackage[usenames,dvipsnames]{color}
\usepackage{mathtools}
\usepackage[normalem]{ulem}

\newcommand{\usenomenclature}{}
\ifdefined\usenomenclature
\usepackage[noprefix]{nomencl}
\fi

\newcommand{\beq}{\begin{equation}}
\newcommand{\eeq}{\end{equation}}
\newcommand{\Eqn}[1] {Eqn.~(\ref{#1})}
\newcommand{\eqn}[1] {eqn.~(\ref{#1})}

\newcommand{\Seccite}[1] {Sec.~\ref{#1}}

\newcommand{\Fig}[1]{Fig.~\ref{#1}}

\newcommand{\mylabel}[1]{\label{#1}} 
\newcommand{\mycite}[1]{\cite{#1}}

\newcommand{\bcen}{\begin{center}}
\newcommand{\ecen}{\end{center}}
\newcommand{\btab}{\begin{tabular}}
\newcommand{\etab}{\end{tabular}}
\newcommand{\bdes}{\begin{description}}
\newcommand{\edes}{\end{description}}
\newcommand{\mc}{\multicolumn}

\newcommand{\bea}{\begin{eqnarray}}
\newcommand{\eea}{\end{eqnarray}}

\newcommand{\bary}{\begin{array}}
\newcommand{\eary}{\end{array}}
\newcommand{\benum}{\begin{enumerate}}
\newcommand{\eenum}{\end{enumerate}}
\newcommand{\bitem}{\begin{itemize}}
\newcommand{\eitem}{\end{itemize}}

\newcommand{\ba} { \bm{a} }

\newcommand{\be} { \mbox{\boldmath $e$}}

\newcommand{\bs} { \mbox{\boldmath $s$}}

\newcommand{\D}[1]{\mbox{d}{#1}}

\newcommand \seccite[1]{sec.~\ref{#1}}

\makeatletter

\newcommand{\Rmnum}[1]{\expandafter\@slowromancap\romannumeral #1@}
\makeatother

\newcommand{\ci}{\mathfrak{i}}

\newcommand{\dtau}{\textup{d}\tau}

\newcommand{\dt}{\textup{d}t}

\newcommand{\cd}{c^\dagger}

\newcommand{\deltau}{\partial_\tau}

\newcommand{\cb}{\bar{c}}

\newcommand{\Tr}{\textup{Tr}}

\newcommand{\tbd}[1]{}

\newcommand{\concept}[1]{}
\newcommand{\figconcept}[1]{}
\newcommand{\eqnconcept}[1]{}
\newcommand{\tabconcept}[1]{}
\newcommand{\appconcept}[1]{}
\newcommand{\appcite}[1]{Appendix \ref{#1}}

\global\long\def\ren{S^{(n)}}

\global\long\def\retwo{S^{(2)}}

\global\long\def\ra{\rangle}

\global\long\def\la{\langle}

\global\long\def\I{\mathbf{1}}

\def\del{\partial}

\usepackage{cancel}

\newcommand{\ee}{\end{align}}

\def \Tr{\mathrm{Tr}}
\def \ra{{\rightarrow}}

\def \be{\begin{equation}}
\def \ee{\end{equation}}
\def \ba{\begin{array}}
\def \ea{\end{array}}
\def \bea{\begin{eqnarray}}
\def \eea{\end{eqnarray}}

\def \a{{\alpha}}

\def \b{{\beta}}

\def \D{{\Delta}}

\def \ra{{\rangle}}
\def \la{{\langle}}
\def \ba{\begin{align*}}
\def \ea{\end{align*}}

\def \bs{\boldsymbol}
\def \mc{\mathcal}

\newcommand {\apgt} {\ {\raise-.5ex\hbox{$\buildrel>\over\sim$}}\ }
\newcommand {\aplt} {\ {\raise-.5ex\hbox{$\buildrel<\over\sim$}}\ }

\def \thop{t_{hop}}
\def \tchain{t_{chain}}
\def\mfp{l_0}
\newcommand {\rem}[1]{}

\newcommand{\response}[1]{#1}

\def \titlename {R\'{e}nyi entanglement entropy of Fermi and non-Fermi liquids: Sachdev-Ye-Kitaev model and dynamical mean field theories}
\def \authornames{Arijit Haldar$^{1,2}$,  Surajit Bera$^1$, and Sumilan Banerjee$^1$}
\def \affiliations{$^1$Centre for Condensed Matter Theory, Department of Physics, Indian Institute 
of Science, Bangalore 560012, India\\
$^2$Department of Physics, University of Toronto, 60 St. George St., Toronto, Ontario, M5S 1A7, Canada}

\begin{document}

\title{\titlename}
\author{\authornames}
\affiliation{\affiliations}
\email{arijit.haldar@utoronto.ca}
\email{sumilan@iisc.ac.in}
\date\today

\begin{abstract}
We present a new method for calculating R\'{e}nyi entanglement entropies for fermionic field-theories originating from microscopic Hamiltonians. The method builds on an operator identity which we discover for the first time. The identity leads to the representation of traces of operator products, and thus R\'{e}nyi entropies of a subsystem, in terms of fermionic-displacement operators. This allows for a very transparent path-integral formulation, both in and out-of-equilibrium, having a simple boundary condition on the fermionic fields. The method is validated by reproducing well known expressions for entanglement entropy in terms of the correlation matrix for non-interacting fermions. We demonstrate the effectiveness of the method by explicitly formulating the field theory for R\'{e}nyi entropy in a few zero and higher-dimensional large-$N$ interacting models akin to the Sachdev-Ye-Kitaev (SYK) model, and for the Hubbard model within the dynamical mean-field theory (DMFT) approximation. We use the formulation to compute R\'{e}nyi entanglement entropy of interacting Fermi liquid (FL) and non-Fermi liquid (NFL) states in the large-$N$ models and compare successfully with the results obtained via exact diagonalization for finite $N$. We elucidate the connection between R\'{e}nyi entanglement entropy and residual entropy of the NFL ground state in the SYK model and extract sharp signatures of quantum phase transition in the entanglement entropy across an NFL to FL transition. Furthermore, we employ the method to obtain nontrivial system-size scaling of entanglement in an interacting diffusive metal described by a chain of SYK dots.
\end{abstract}

\maketitle
\section{Introduction}
Quantum entanglement has emerged as a major tool to characterize quantum phases and phase transitions \cite{Calabrese2004,Calabrese2009,Casini2009,Eisert2010,Refael2009,Laflorencie2016} and to distill fundamental quantum mechanical nature of non-trivial many-body states, e.g. those with topological order that is otherwise hard to quantify \cite{Kitaev2006,Jiang2012}. Recent developments in condensed matter and high energy physics have revealed beautiful connections between entanglement, thermalization and dynamics, leading to classification of dynamical phases of interacting quantum systems into thermal and many-body localized (MBL) phases \cite{Nandkishore2015,Altman2015,Abanin2019}. Quantum entanglement is quantified in terms of properties of the reduced density matrix, $\rho_A=\mathrm{Tr}_B\rho$, of a system with density matrix $\rho$ and divided, e.g., into subsystems $A$ and $B$, where $\mathrm{Tr}_B$ is the partial trace over the degrees of freedom of $B$. Typical entanglement measures constructed out of $\rho_A$ for a pure state are von Neumann and R\'{e}nyi entanglement entropies. The latter can be used to compute useful measures \cite{Vidal2002,Calabrese2012,Lee2013,Wu2019,Lu2019} for entanglement even in a thermal mixed state. 

However, calculation of entanglement entropy is much more challenging than, e.g. that of thermal entropy or usual correlation functions. 
Over the last decade a lot of progress has been made to obtain entanglement entropy, both numerically and analytically, for non-interacting bosonic and fermionic systems \cite{Vidal2003,Casini2009,Eisert2010}, and at critical points described by conformal field theories \cite{Calabrese2004,Calabrese2009}.
The latter rely on field-theoretic techniques using replicas and path integrals, typically in imaginary time, with complicated boundary conditions on fields and associated Green's function along the time direction \cite{Calabrese2004,Calabrese2009,Casini2009}. Such methods are often hard to implement for interacting systems. Hence, computation of entanglement entropy for interacting systems are only limited to small systems using exact diagonalization (ED) or for systems, like those without sign problem accessible via quantum Monte Carlo (QMC) simulations \cite{Hastings2010,Humeniuk2012,Grover2013,Wang2014}. 

A promising path-integral approach, that circumvents the use of complicated boundary condition using known relation between reduced density matrix and Wigner characteristic function \cite{Cahill1969}, has been recently proposed for bosonic systems in ref.\onlinecite{Sensarma2018}. Motivated by this, here we develop a new field theoretic method to compute R\'{e}nyi entanglement entropy for fermions. A similar field theory formalism for fermions has been developed independently by Moitra and Sensarma \cite{Moitra2020}. To this end, we derive new representations of an operator and traces of product of operators in terms of fermionic displacement operators \mycite{Cahill1999}. The representations allow us to develop a very transparent fermionic coherent-state path-integral method with simple boundary condition to compute R\'{e}nyi entropies of a sub-region of the system in terms of a fermionic version of the `Wigner characteristic function'. The formalism can equally be applied to calculate subsystem R\'{e}nyi entropy for thermal equilibrium state via imaginary-time path integral or non-equilibrium time evolution described via Schwinger-Keldysh field theory. The approach naturally transcends the effect of boundary condition into time-dependent self-energy, which acts like `kick' at a particular time. We show that the method immediately reproduces the known expressions for von Neumann and R\'{e}nyi entanglement entropies for non-interacting fermions. The effect of the time-dependent self-energy can be implemented for interacting systems treated within standard perturbative and non-perturbative field-theoretic approximation and diagrammatic continuous-time Monte Carlo simulation \cite{Gull2011}. We elucidate this by deriving the subsystem second R\'{e}nyi enetropy within two well-known approaches to treat correlated fermions - (a) strongly interacting large-$N$ fermionic models based on Sachdev-Ye-Kitaev model \cite{KitaevKITP,Sachdev1993}, and (b) dynamical mean-field theory (DMFT) \cite{Georges1996RMP}.

In the other major part of the paper we explicitly demonstrate the utility of the method by computing the second R\'{e}nyi entropy ($S^{(2)}$) for subsystems in several large-$N$ model in thermal equilibrium - (i) zero-dimensional SYK model having infinite-range random four-fermion or two-body interaction with a non-Fermi liquid (NFL) ground state, (ii) SYK model with additional quadratic hopping between fermions having a Fermi liquid (FL) ground state, (iii) a generalized SYK model, the Banerjee-Altman (BA) model \cite{BanerjeeAltman2016}, having quantum phase transition (QPT) between SYK NFL and FL, and (iv) an extended system, a chain of SYK dots \cite{Gu2016,Song2017}, describing an interacting diffusive metal. We compute the subsystem R\'{e}nyi entropy at the large-$N$ saddle point as a function of subsystem size for the thermal density matrix at a temperature $T$ and extrapolate to $T\to 0$ to obtain ground-state R\'{e}nyi entanglement entropy. In all the above cases except (iii), we compare and contrast the results for the interacting systems with a corresponding non-interacting system where the SYK interaction is replaced by infinite-range random hopping. For the non-interacting models we numerically calculate the second R\'{e}nyi entropy by numerical diagonalization for moderately large systems and compare with large-$N$ field theoretic results. Moreover, we also compute subsystem R\'{e}nyi entropy via exact diagonalization (ED) of many-body Hamiltonian for small $N\sim 8-16$ for the interacting zero-dimensional models. We obtain the following important results using our method for the large-$N$ models.

(1) We show that for the SYK model, in the $N\to\infty$ limit, the zero-temperature residual entropy \cite{Sachdev2015,KitaevKITP,Kitaev2018,Maldacena2016} of the SYK NFL contributes to the $T=0$ subsystem R\'{e}nyi entropy, thus making it difficult to recover the true quantum entanglement of the NFL ground state starting from a thermal ensemble. Moreover, consistent with our numerical results, we analytically prove that the SYK model is maximally entangled when the relative size of the subsystem $p\to 0$.

(2) We demonstrate how the $T\to 0$ bipartite R\'{e}nyi entanglement entropy crosses over from a NFL to a FL as function of the strength of hopping in the SYK model with random quadratic term. The results show that heavy FL are much more entangled than weakly- or non-interacting FL.

(3) In the BA model with a NFL-FL transition, we establish a precise connection between subsystem R\'{e}nyi entropy and the residual entropy of the NFL, and show that the R\'{e}nyi entropy also carries sharp signature of the underlying QPT, like the residual entropy \cite{BanerjeeAltman2016}. 

(4) In the extended one-dimensional (1D) model of SYK dots, we compute R\'{e}nyi entanglement entropy of a subsystem of length $l$. We find a crossover from $S^{(2)}\sim \log l$ to $S^{(2)}\sim \log(1/(l^{-2}+\mfp^{-2})^{1/2})$ with increasing $l$ where the $\log l$ behavior, expected for gapless fermions, gets cut off by an emergent mean-free path $\mfp$ in an interacting diffusive metal.

The paper is organized as follows. In \Seccite{sec:formulation} we derive the operator identities that form the basis of our formalism and discuss the connections of these identities with R\'{e}nyi entropy of a subsystem. The general formulation for the equilibrium and non-equilibrium path integrals to compute the R\'{e}nyi entropy is discussed in \Seccite{subsec:RenyiFieldTheory}. Sec. \ref{subsec:nI-formalism} describes the application of the field-theory formalism to derive well-known formulae for R\'{e}nyi entropy of non-interacting fermions. In Secs. \ref{sec:allmodels} and \ref{subsec:DMFT}, we develop the field-theory for R\'{e}nyi entropy in several interacting large-$N$ models based on the SYK model, and in the Hubbard model within DMFT approximation, respectively. We describe our analytical and numerical results for R\'{e}nyi entropy in the large-$N$ models and the comparison of the large-$N$ results with that obtained from numerical exact diagonalization in \Seccite{sec:Results}. Additional details of the derivations of the operator identities, path integral formulations and their analytical and numerical implementations in various models for computing R\'{e}nyi entropies are given in the appendices.   

\section{Coherent state-path integral formalism for fermions in and out-of equilibrium}
\subsection{Subsystem R\'{e}nyi entropy, displacement operator and trace formula}\mylabel{sec:formulation}
In this section, we consider a system with fermionic degrees of freedom and derive a useful expansion of an arbitrary operator in terms of the so-called fermionic displacement operator \cite{Cahill1999} and show that the expansion can be used to represent R\'{e}nyi entropies of a sub-region of the system. A similar representation of R\'{e}nyi entropy has been independently developed by Moitra and Sensarma \cite{Moitra2020}. The R\'{e}nyi  entropy of a quantum system described by a density matrix $\rho$ is obtained by dividing the system into two parts $A$ and $B$ (not necessarily equal) and defining a \emph{reduced density matrix}, 
\begin{align}\mylabel{eq:red-dmat-def}
    \rho_A=\Tr_B\rho,
\end{align}
for region $A$. If $\rho$ represents a pure state, then a measure of the quantum entanglement of region $A$ with $B$ can be obtained by evaluating the von Neumann entropy $S_A=-\Tr_A[\rho_A\ln \rho_A]$ by tracing over the degrees of freedom in $A$. In practice however, the von Neumann entropy is often hard to calculate directly within field theoretic methods, and a more convenient measure of entanglement \mycite{Calabrese2004,Calabrese2009,Casini2009,Eisert2010} is the $n$-th R\'{e}nyi entropy, $S_A^{(n)}$, defined as
\begin{align}\mylabel{eq:n-Renyi-def}
    \ren_A=\frac{1}{1-n}\log\Tr_{A}\left[\rho_{A}^{n}\right]
\end{align}
where the integer $n>1$. The von Neumann entanglement entropy can be obtained by analytically continuing to $n\to 1$. We refer to the above subsystem R\'{e}nyi entropy throughout the paper simply as \emph{R\'{e}nyi entropy} for brevity. 

The main difficulty in evaluating the above comes from the representation of  $\Tr_A[\rho_A^n]=\Tr_A[(\Tr_B \rho)(\Tr_B \rho)\dots(\Tr_B \rho)]$ in a coherent-state path integral, since each factor of $\Tr_B\rho$ leads to separate replicas which need to be connected via appropriate boundary condition when represented through Grassmann variables \cite{Calabrese2004,Casini2009}. To circumvent this difficulty, we derive the following operator expansion (see \appcite{app:OperatorExpansion}) for an arbitrary operator $F$,
\begin{align}\mylabel{eq:op-exp-Dn-Dn}
    F=\int d^{2}(\bs{\xi},\bs{\eta})\ f_{N}(\bs{\eta},\bs{\xi})\Tr[FD_{N}(\bs{\xi})]D_{N}(\bs{\eta}),
\end{align}
where $\bs{\xi}\equiv \{\bar{\xi}_i,\xi_i\}$ and $\bs{\eta}\equiv \{\bar{\eta}_i,\eta_i\}$ denote set of Grassmann variables with index $i=1,\dots,N_A$, e.g., referring to a set of sites that includes the support of the operator $F$ on the lattice; $d^2(\bs{\xi},\bs{\eta})=\prod_id\bar{\xi}_id\xi_id\bar{\eta}_id\eta_i$. Here the weight function $f_N(\bs{\xi},\bs{\eta})$ is given by
\begin{align}\mylabel{eq:fn-gamma-xi-def}
    f_{N}(\bs{\eta},\bs{\xi})=&2^{N_A}\exp\left[{-\frac{1}{2}\sum_{i}(\bar{\eta}_i\eta_{i}+\bar{\xi}_{i}\xi_{i}-\bar{\eta}_i\xi_{i}+\bar{\xi}_{i}\eta_{i})}\right].
\end{align}
The basis of above expansion in \Eqn{eq:op-exp-Dn-Dn} is formed by the fermionic \emph{displacement operators} \mycite{Cahill1999}, much like more familiar bosonic counterparts \cite{Cahill1969}, defined as
\begin{align}\mylabel{eq:Dop-def-0}
    D(\bs{\xi})=\exp\left[\sum_{i}(\cd_{i}\xi_{i}-\bar{\xi_{i}}c_{i})\right],
\end{align}
where $\cd_i$, $c_i$ are the creation and annihilation operators on site $i$. The displacement operator generates the coherent state \cite{Cahill1999}, $|\xi\rangle=D(\bs{\xi})|0\rangle$ by shifting the vacuum state $|0\rangle$ such that $c_i|\xi\rangle=\xi_i|\xi\rangle$. The displacement operator $D_N(\bs{\xi})$ that appears in \Eqn{eq:op-exp-Dn-Dn} is normal ordered,
\begin{align}\mylabel{eq:Dn-def}
    D_{N}(\bs{\xi})=&\exp\left[{\sum_{i}\cd_{i}\xi_{i}}\right]\exp\left[{-\sum_{i}\bar{\xi_{i}}c_{i}}\right]\notag\\
    =&D(\bs{\xi})\exp\left[{\frac{1}{2}\sum_{i}\bar{\xi_{i}}\xi_{i}}\right],
\end{align}
and more convenient to be used in the path integral representation discussed in the next section.

\Eqn{eq:op-exp-Dn-Dn} offers a way to decompose a general operator $F$ using \emph{only} the normal-ordered displacement operators $D_N(\bs{\xi})$ and is one of the \emph{key} results of this paper. An important corollary to the decomposition identity of \Eqn{eq:op-exp-Dn-Dn} is to express the trace of the product of operators $F$ and $G$ as
\begin{align}\mylabel{eq:Tr-FG-exp-DN}
    &\Tr\left[FG\right]=\int d^{2}(\bs{\xi},\bs{\eta}) f_{N}(\bs{\xi},\bs{\eta})\Tr[FD_{N}(\bs{\xi})]\Tr[GD_{N}(\bs{\eta})].
\end{align}
To derive the above, we have used the identities, $\Tr[D_N(\bs{\eta})G]=\Tr[GD_N(-\bs{\eta})]$ and $f_N(-\bs{\eta},\bs{\xi})=f_N(\bs{\xi},\bs{\eta})$ (see \appcite{app:OperatorExpansion}).
\Eqn{eq:Tr-FG-exp-DN} is the second \emph{key} result of this paper and is crucial for deriving a path-integral representation for evaluating \response{R\'{e}nyi entropy} as we discuss in the next section.
We could also make the operator expansion, $F=\int d^2(\bs{\xi},\bs{\eta})f(\bs{\eta},\bs{\xi})\Tr[FD(\bs{\xi})]D(\bs{\eta})$ and obtain the corresponding trace formula similar to \Eqn{eq:Tr-FG-exp-DN}, in terms of the displacement operator $D(\bs{\xi})$ and weight function $f(\bs{\eta},\bs{\xi})=2^{N_A}\exp(\sum_i(\bar{\eta_i}\xi_i-\bar{\xi}_i\eta_i)/2)$.   

Using the trace formula (\Eqn{eq:Tr-FG-exp-DN}), e.g., the second-R\'{e}nyi entropy, $S^{(2)}_A$, can be conveniently expressed as
\begin{align}\mylabel{eq:expS2-expression-red}
    e^{-\retwo_A}=&\Tr_{A}\left[\rho_{A}\rho_{A}\right]\notag\\
	=&\int d^{2}(\bs{\xi},\bs{\eta}) f_N(\bs{\xi},\bs{\eta})\Tr_{A}[\rho_{A}D_N(\bs{\xi})]\Tr_{A}[\rho_{A}D_N(\bs{\eta})].
\end{align}
In the above, $\bs{\xi}\equiv\{\bar{\xi}_i,\xi_{i}\}_{i\in A}$ (and similarly for $\bs{\eta}$), i.e. the displacements operators above only involve the fermionic operators in the region $A$. 
As a result, we have
\begin{align}\mylabel{eq:reduced-to-full}
    \Tr_{A}[\rho_{A}D_N(\bs{\xi})]=\Tr[\rho D_N(\bs{\xi}\in A)],
\end{align}
i.e., the expectation value of the operator $D(\bs{\xi})$ evaluated for the reduced system is same as that obtained using the \emph{full} density matrix. \Eqn{eq:reduced-to-full}, therefore, eliminates the need to calculate the reduced density matrix $\rho_A$. The evaluation of $\rho_A$ is the difficult step for calculating entanglement entropy, as mentioned earlier. 
To proceed further, we define the `normal-ordered' \emph{fermionic Wigner characteristic function} \cite{Cahill1969,Cahill1999} for the density matrix,
\begin{align}\mylabel{eq:char-func}
\chi_{N}(\bs{\xi})	&=\Tr[\rho D_{N}(\bs{\xi}\in A)],
\end{align}
and arrive at the final expression for the second-R\'{e}nyi entropy for region $A$
\begin{align}\mylabel{eq:expS2-expression}
    e^{-\retwo_A}=&\int_{\bs{\xi},\bs{\eta}\in A} d^{2}(\bs{\xi},\bs{\eta}) f_N(\bs{\xi},\bs{\eta})\chi_N(\bs{\xi})\chi_N(\bs{\eta}).
\end{align}
We also sometime use an analogous expression written in terms of $D(\bs{\xi})$ by replacing $D_N(\bs{\xi})$ and the function $f_N(\bs{\xi},\bs{\eta})$ by $f(\bs{\xi},\bs{\eta})$. This leads to the usual fermionic characteristic function \cite{Cahill1999},
\begin{align}
\chi(\bs{\xi})&=\Tr[\rho D(\bs{\xi}\in A)]. \label{eq:char-func_1}
\end{align}
The higher-order R\'{e}nyi entropies,  $S^{(n>2)}_A$, can be found in a similar manner by repeated application of the trace identity in \Eqn{eq:Tr-FG-exp-DN}. In fact, as we show in \Seccite{subsec:nI-formalism}, a hierarchy for higher-order R\'{e}nyi entropies can be derived that recursively expresses the characteristic function of higher moments of the reduced density matrix $\rho_A$ in terms of the lower order ones.

\subsection{Equilibrium and non-equilibrium field theories for R\'{e}nyi entropy} \label{subsec:RenyiFieldTheory}
A path integral representation of the characteristic function, $\chi_N(\bs{\xi})$, will naturally lead to a similar representation for the second R\'{e}nyi entropy of region $A$ through \Eqn{eq:expS2-expression}. Therefore, we first derive the path integral for $\chi_N(\bs{\xi})$ for -- (a) thermal density matrix describing a system in equilibrium, and (b) time-evolving density matrix subjected to a general time-dependent Hamiltonian, e.g. describing a quench.

\subsubsection{Path integral for thermal equilibrium}\mylabel{para:Equilibrium}
The density matrix for a system described by a Hamiltonian $H$ under thermal equilibrium is given by
\begin{align}\mylabel{eq:rho-therm}
    \rho=Z^{-1}\exp[-\beta H],
\end{align}
with the partition function 
\begin{align}\mylabel{eq:Z-def}
Z=\Tr\left[\exp\left(-\beta H\right)\right]    
\end{align} and inverse temperature $\beta=1/T$ ($k_\mathrm{B}=1$). 
 Inserting the identity operator $\int d^2c|c\rangle \langle c|=I$ in the coherent state basis, we get
\begin{align}
\chi_N(\bs{\xi})=&Z^{-1}\int\prod_{n=0}^{1}d^{2}c_n \la-c_0|e^{-\beta H}|c_1\ra\la c_{1}|D_{N}(\bs{\xi})|c_{0}\ra,
\end{align}
with $d^2c_n=\prod_i d\bar{c}_{in}dc_{in}$.
Using \Eqn{eq:Dn-def}, the matrix element for the displacement operator is easily evaluated as
\begin{align}\mylabel{eqn:Dinp}
\la c_{1}|D_{N}(\bs{\xi})|c_{0}\ra=\exp\left(\sum_{i\in A}\cb_{i,1}\xi_i-\bar{\xi}_ic_{i,0}\right)
\la c_1|c_0\ra.    
\end{align}
 Finally, following the standard methodology \cite{AltlandSimonsBook} of fermionic coherent state, e.g., evaluating $\la -c_0|e^{-\beta H}|c_1\ra$ via Trotter decomposition $\beta=N_{\tau}\Delta\tau$ and taking the continuum limit $N_\tau\to\infty,\Delta \tau\to 0$ we get
\begin{widetext}
\begin{align}\mylabel{eq:chi-path-int}
    \chi_N(\bs{\xi})=Z^{-1}\int \mathcal{D}(\bar{c},c)
    \exp\left[-\int_0^\beta d\tau\left\{ \sum_i\cb_i(\tau)\partial_{\tau}c_i(\tau)+H(\cb_i(\tau),c_i(\tau))
    -\sum_{i\in A}\cb_i(\tau)\delta(\tau^{+})\xi_i+\sum_{i\in A}\bar{\xi}_i\delta(\tau)c_i(\tau)\right\} \right],
\end{align}
\end{widetext}
as the path-integral representation of the characteristic function. The above equation offers a significant advantage over the usual field-theoretic formalism \mycite{Calabrese2004,Calabrese2009,Casini2009} used to compute \response{R\'{e}nyi entropy}. The imaginary-time boundary condition for the Grassmann fields $c_i(\tau)$ and $\cb_i(\tau)$ are still antiperiodic, i.e. $c_i(\tau+\beta)=-c_i(\tau)$, irrespective of whether $i$ belongs to the region $A$ or not. Instead, the distinction between subsystem $A$ and the rest of the system is encoded by the auxiliary fields $\bs{\xi}$ which only couple with the fields $c_i(\tau)$ and $\cb_i(\tau)$ for $i\in A$. Under time discretization, relevant for the numerical implementation discussed later, we have
\begin{align}\mylabel{eqn:finite-diff-def}
   \cb_i(\tau)\partial_{\tau}c_i(\tau)=&\cb_{i,n}\frac{(c_{i,n}-c_{i,n-1})}{\D\tau}\notag\\
    \delta(\tau^{+})	=&\frac{1}{\D\tau}\delta_{n,1},\ \ \ \ 
   	\delta(\tau)=\frac{1}{\D\tau}\delta_{0,n},
\end{align}
where $n$ denotes the $n$-th time index.

\subsubsection{Path integral for non-equilibrium evolution}
\begin{figure}
    \centering
    \includegraphics[width=0.48\textwidth]{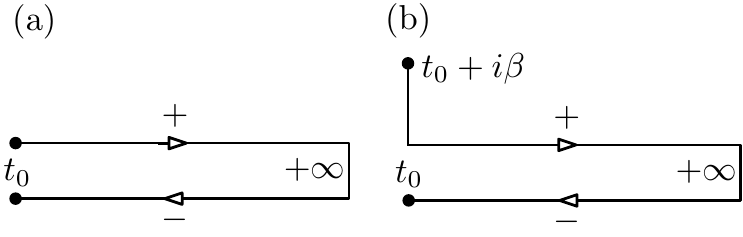}
    \caption{Closed time contours (CTCs): (a) The Schwinger-Keldysh contour, for an arbitrary initial density matrix $\rho_0$, starting at time $t_0$ extending to $+\infty$ and returning back to $t_0$. The two branches represent the forward (+) and backward (-) evolution of time respectively. (b) The modified contour for an initial density matrix picked from a thermal ensemble, i.e., $\rho_0\sim\exp(-\beta H)$. The $\exp(-\beta H)$ term is incorporated into the contour as an evolution in imaginary time, represented by the additional vertical branch of length $\beta$.}
    \label{fig:CTCs}
\end{figure}

\mylabel{para:Keldysh-chi}
The density matrix, at a time $t$, evolving under a time-dependent Hamiltonian $H(t)$ is given by
\begin{align}
   \rho(t)=U(t,t_0)\rho_0 U(t_0,t), 
\end{align}
where $\rho_0$ is the initial density matrix at time $t_0$. The operator $U(t_1,t_2)$ is the unitary evolution operator associated with the Hamiltonian $H(t)$ and defined as
\begin{align}\mylabel{eqn:U-def}
    U(t_1,t_2)=
    \begin{cases}
    \textup{T}\left[\exp\left(-\ci\int_{t_1}^{t_2} \dt\ H(t)\right)\right] & t_1\geq t_2\\
    \tilde{\textup{T}}\left[\exp\left(-\ci\int_{t_1}^{t_2} \dt\ H(t)\right)\right]& t_1< t_2
    \end{cases},
\end{align}
where $\textup{T}$, $\tilde{\textup{T}}$ are the time ordering and anti-time ordering operators respectively. We use Schwinger-Keldysh closed time-contour formalism \mycite{KamenevBook,StefanucciBook} to obtain a path integral representation for the R\'{e}nyi entropy, e.g., $S^{(2)}(t)$, which is now time dependent and given by
\begin{align}\mylabel{eq:expS2-expression_t}
    e^{-\retwo_A(t)}=&\int_{\bs{\xi},\bs{\eta}\in A} d^{2}(\bs{\xi},\bs{\eta}) f_N(\bs{\xi},\bs{\eta})\chi_N(\bs{\xi},t)\chi_N(\bs{\eta},t).
\end{align}
Here we have rewritten the trace identity (\Eqn{eq:Tr-FG-exp-DN}) (see \appcite{app:OperatorExpansion}) to define the time-dependent characteristic function as
\begin{align}\mylabel{eq:chi-neq-def}
\chi_N(\bs{\xi},t)=&\Tr[D_{N}(\bs{\xi})\rho(t)]=\Tr[U(t_0,t)D_{N}(\bs{\xi})U(t,t_0)\rho_0],
\end{align}
for the sake of convenience in constructing the path integral representation.
As in the standard Schwinger-Keldysh closed-time contour formalism \cite{KamenevBook,StefanucciBook}, the last line in \Eqn{eq:chi-neq-def} may be interpreted, from right to left, as starting from an initial density matrix $\rho_0$, evolving forward in time (represented by $+$ branch in \Fig{fig:CTCs}(a)) by $U(t,t_0)$ from $t_0$ to $t$, applying  fermionic source fields $\bs{\xi}$ in region $A$, through $D_N(\bs{\xi})$, at time $t$ and then going back to $t_0$ via the backward time-evolution ($-$ branch in \Fig{fig:CTCs}(a)) $U(t_0,t)$. 
We refer to the closed-time contour with the symbol $\mathcal{C}$ and use a contour variable $z$, which takes values $(t,+)$, $(t,-)$ at time $t$ for the $+$, $-$ branches. As done often in the Schwinger-Keldysh formalism, the contour is extended to $+\infty$ (see \Fig{fig:CTCs}). We incorporate this contour extension in our expression for the characteristic function. 
\begin{align}\mylabel{eq:chi-neq-def-2} 
\chi_N(\bs{\xi},t)=&\Tr[U(t_0,t)U(t,\infty)U(\infty,t)D_{N}(\bs{\xi})U(t,t_0)\rho_0].    
\end{align}

Again following standard route \cite{AltlandSimonsBook,KamenevBook}, as in the thermal equilibrium case, we obtain a path integral representation for $\chi_N(\bs{\xi},t)$, in terms of the \emph{entanglement Keldysh action}

\begin{align}\mylabel{eqn:SEEK-def}
&\mathcal{S}^\mathcal{C} =\int_{\mathcal{C}}dz\left[\sum_{i}\bar{c}_{i}(z)\ci\partial_{z}c_{i}(z)
 -H(\bar{c}(z),c(z))\right]\notag\\
 &-\ci\int_{\mathcal{C}}dz\sum_{i\in A}\left[\bar{c}_{i}(z)\delta_\mathcal{C}\left(z,(t^+,+)\right)\xi_i-\bar{\xi}_i\delta_\mathcal{C}\left(z,(t,+)\right)c_{i}(z)\right],   
\end{align}
such that
\begin{align}\mylabel{eqn:chi-Keldysh}
    \chi_N(\bs{\xi},t)=&\int\mathcal{D}(\bar{c},c)e^{i\mathcal{S}^\mathcal{C}}
    \langle c_i(0,+)|\rho_{0}|-c_i(0,-)\rangle.
\end{align}
Here $\delta_\mathcal{C}(z,z_1)$ is the Dirac delta function on $\mathcal{C}$ and takes into account both the time argument and the branch index ($\pm$).
The coherent-state matrix element of the initial density matrix, $\langle c_i(0,+)|\rho_{0}|-c_i(0,-)\rangle$ in \Eqn{eqn:chi-Keldysh} encodes the information of the initial state at starting time $t_0$.
For a thermal initial state, one can add an additional branch to $\mathcal{C}$ and redefine it as 
\begin{align}\mylabel{eqn:CTC1-def}
    \mathcal{C}=[t_0+\ci\beta,t_0)\cup [t_0,+\infty)\cup(+\infty,t_0],
\end{align}
i.e., the contour is now comprised of a vertical imaginary-time branch of length $\beta$ and the usual $+$ and $-$ branches, as shown in \Fig{fig:CTCs}(b). The form of the action $\mathcal{S}^\mathcal{C}$ remains exactly the same as given in \Eqn{eqn:SEEK-def} such that
\begin{align}\mylabel{eqn:chi-Keldysh-thermal}
    \chi_N(\bs{\xi,t})=&Z_0^{-1}\int\mathcal{D}(\bar{c},c)e^{i\mathcal{S}^\mathcal{C}},
\end{align}
where $Z_0$ is the partition function for the initial thermal state described by $\rho_0$.
We emphasize here that the Hamiltonian describing the $\rho_0$ and the one dictating the unitary evolution (see \Eqn{eqn:U-def}) \emph{do not} need to be same, e.g. in a quench \cite{Moeckel2009,Dziarmaga2010,Polkovnikov2011,Essler2016}. This distinction is incorporated by choosing the appropriate Hamiltonian in \Eqn{eqn:SEEK-def} for the imaginary-time and real-time branches.

In the next section we show that the equilibrium and non-equilibrium coherent-state path integrals can be used immediately to obtain well-known expressions for R\'{e}nyi and von Neumann entanglement entropies in terms of the correlation matrix \cite{Casini2009}.


\subsection{R\'{e}nyi entropies for non-interacting fermions}\mylabel{subsec:nI-formalism}
We consider a non-interacting system described by the thermal density matrix
\beq
\rho=\exp\left[-\beta \sum_{ij}t_{ij}\cd_ic_j\right]/Z , \label{eq:ThermalNonInt}
\eeq
where $Z=\Tr\exp{\left[-\beta \sum_{ij}t_{ij}\cd_ic_j\right]}$ is the partition function. 
Using \Eqn{eq:chi-path-int}, the normal-ordered characteristic function in this case is given by
\begin{align}
\chi_{N}(\bs{\xi})	=&Z^{-1}\int\mathcal{D}(\bar{c},c)\notag\\
&\exp\left[-\int_{0}^{\beta}d\tau\left\{ \sum_{ij}\bar{c_{i}}(\tau)(\delta_{ij}\partial_{\tau}+t_{ij})c_{j}(\tau)\right.\right.\notag\\
&\qquad\left.\left.-\sum_{i\in A}\bar{c_{i}}(\tau)\delta(\tau^{+})\xi_{i}+\sum_{i\in A}\bar{\xi_{i}}\delta(\tau)c_{i}(\tau)\right\} \right].
\end{align}
The Gaussian structure of the fermionic-integral allows us to integrate the fermions 
to give
\begin{align}\mylabel{eq:chiN-nI}
\chi_{N}(\bs{\xi})
=	
&\exp\left[-\int\dtau_{1,2}\sum_{ij}\bar{\xi_{i}}\delta(\tau_{1})(-G)_{ij}(\tau_{1},\tau_{2})\delta(\tau_{2}^{+})\xi_{j}\right],    
\end{align}
where the matrix $G$, is the Green's function and is defined as the imaginary-time-ordered two point correlator
\begin{align}\mylabel{eq:thermal-G-def}
    G_{ij}(\tau_{1},\tau_{2})=&-\la T_{\tau}c_{i}(\tau_{1})\cd_{j}(\tau_{2})\ra\notag\\=&\begin{cases}
-\la c_{i}(\tau_{1})\cd_{j}(\tau_{2})\ra & \tau_{1}>\tau_{2}\\
+\la\cd_{j}(\tau_{2})c_{i}(\tau_{1})\ra & \tau_{1}<\tau_{2},
\end{cases}
\end{align}
and can be calculated using 
\begin{align}
(-G^{-1})_{ij}(\tau_{1},\tau_{2})=(\delta_{ij}\partial_{\tau}+t_{ij})\delta(\tau_{1}-\tau_{2}).    
\end{align}
However, because of the delta functions in \Eqn{eq:chiN-nI}, we need to evaluate only $G_{ij}(0,0^{+})$. This allows us to define a \emph{correlation matrix} \mycite{Casini2009} $C_{ij}$, as follows
\begin{align}\mylabel{eq:corr-func-def}
 C^T_{ij}=C_{ji}\equiv G_{ij}(0,0^{+})=\la\cd_{j}c_{i}\ra=\Tr\left[\rho\cd_{j}c_{i}\right],
\end{align}
where $C^T$ is transpose of the matrix $C$. Therefore, the characteristic functions for an arbitrary non-interacting system is given by
\begin{align}
\chi_{N}(\bs{\xi})	&=\exp\left[\sum_{ij\in A}\bar{\xi_{i}}C^T_{ij}\xi_{j}\right]\mylabel{eq:chi-nI-final-ans}\\
\chi(\bs{\xi})	&=\exp\left[\sum_{ij\in A}\bar{\xi_{i}}C^T_{ij}\xi_{j}-\frac{1}{2}\sum_{i\in A}\bar{\xi_{i}}\xi_{i}\right]\mylabel{eq:chi-nI-final-ans}.
\end{align}
The Gaussian structure of the characteristic function is a direct consequence of the underlying non-interacting Hamiltonian. Using \Eqn{eq:expS2-expression} and the characteristic function above, the second R\'{e}nyi entropy can be immediately evaluated by integrating out the auxiliary Grassmann variables $\bs{\xi}$ and $\bs{\eta}$ to get the well-known expression \cite{Casini2009} for non-interacting fermions, 
\begin{align}\mylabel{eqn:S2-nI-thermal}
    S^{(2)}	
	&=-\Tr\ln\left[(\I-C)^{2}+C^{2}\right],
\end{align}
where $\I$ is a $N_A\times N_A$ identity matrix and only the elements $C_{ij}$ of the correlation matrix (see \Eqn{eq:corr-func-def}) with $i,j\in A$ are involved in the above expression. 
In fact, the Gaussian form of the characteristic functions allow us to derive the expression for the higher R\'{e}nyi entropies as well. To this end, we first define the higher-order generalization of the characteristic function (see \Eqn{eq:char-func})
\begin{align}
    \chi_n(\bs{\xi})=\Tr_{A}[\rho_{A}^{n}D(\bs{\xi}\in A)].
\end{align}
A recursion relation of the form
\begin{align}\mylabel{eq:recurse_1}
\chi_{n+1}(\bs{\alpha})=\frac{1}{2^{N_{A}}}\int d^{2}(\bs{\xi},\bs{\eta})\ f(\bs{\xi},\bs{\eta})f(\bs{\eta},\bs{\alpha}) \chi_{n}(\bs{\xi})\chi_{1}(\bs{\alpha}+\bs{\eta}),    
\end{align}
with $\chi_{n=1}=\chi$, can then be derived (see \appcite{part:Deriving-hiearchy-details}), which obtains higher-order $\chi$ from lower ones. As alluded earlier, we use a Gaussian-ansatz of the form
\begin{align}\mylabel{eq:ch-nI-gaussian-ansatz}
\chi_{n}(\bs{\alpha})=\lambda_{n}\exp\left[\bar{\bm{\alpha}}^{T}\bm{A_{n}}\bm{\alpha}\right],    
\end{align}
where $\bm{\alpha}=\left[\begin{array}{ccc}\alpha_{1} & \cdots & \alpha_{N_{A}}\end{array}\right]^T$ and $\bm{A_{n}}$ is a $N_{A}\times N_{A}$ complex matrix, to solve the recursion. The ansatz is consistent with the expression for $\chi$ obtained in \Eqn{eq:chi-nI-final-ans}, provided we set 
\begin{align}
    \bm{A_{1}}	=&\bm{A}=\bm{C}^T-\I/2\\
\lambda_{1}	=&\lambda=1.
\end{align}
 The ansatz lets us simplify the recursion, see \appcite{subsec:Simplification-of-the-RE-recursion}, and derive recursions involving $\lambda_{n}$ and $A_n$, i.e. 
\begin{align}
    \lambda_{n+1}	=	\det[2\bm{A_{n}A}+\I/2]\lambda_{n}\notag\\
    \bm{A_{n+1}}	=	(\bm{A+A_{n}})(\bm{\I+4AA_{n}})^{-1}.\mylabel{eq:recursion_An_A-main}
\end{align}
The last equation can be rearranged (see \appcite{subsec:Simplification-of-the-RE-recursion}) into
\begin{align}\mylabel{eq:final_X_recursion}
    \bm{X_{n+1}}=\I+\bm{C}^T(\bm{C}^T-\I)\bm{X_{n}}^{-1},
\end{align}
with $X_n$ defined as 
\begin{align}\mylabel{eq:nI-Xn-def}
    \bm{X_{n}}=2A_{n}A+\frac{\I}{2}.
\end{align}
We find the following 
\begin{align}\mylabel{eq:Xn_rec_sol}
    \bm{X_{n}}=\left[(\I-\bm{C}^T)^{n}+\bm{(C^T)}^{n}\right]\left[(\I-\bm{C^T})^{n-1}+\bm{(C^T)}^{n-1}\right]^{-1},
\end{align}
solves \Eqn{eq:final_X_recursion}, see \appcite{sec:sol-verification}. The $n$-th R\'{e}nyi entropy is obtained by evaluating $\lambda_{n}$, since
\begin{align}
    \chi_{n}(\bs{\alpha}=0)=\Tr_{A}\left[\left(\rho_{A}\right)^{n}D(\bs{\alpha}=0)\right]=\Tr_{A}\left[\left(\rho_{A}\right)^{n}\right]=\lambda_{n}, 
\end{align}
and \Eqn{eq:n-Renyi-def} implies
\begin{align}
    \ren_A=\frac{1}{1-n}\log\Tr_{A}\left[\left(\rho_{A}\right)^{n}\right]=\frac{1}{1-n}\log\lambda_{n}.
\end{align}
From the recursion for $\lambda_{n}$ in \Eqn{eq:recursion_An_A-main}, and \Eqn{eq:nI-Xn-def}, we see that
\begin{align}
    \lambda_{n}=\det(\bm{X}_{n})\lambda_{n-1}=\det(X_{n}X_{n-1}\cdots X_{1}).
\end{align}
Looking at the structure of $X_{n}$ in \Eqn{eq:Xn_rec_sol}, we find that the inverse in $X_{n}$ cancels with the non-inverse term in $X_{n-1}$, so on and so forth, leaving at the end
\begin{align}
    \Tr_{A}[\rho_{A}^{n}]=\lambda_{n}&=\det[\left(\I-\bm{C^T}\right)^{n}+\bm{(C^T)}^{n}]\notag\\
    &=\det[(\I-\bm{C})^{n}+\bm{C}^{n}],
\end{align}
from which we find the $n$-th R\'{e}nyi entropy to be
\begin{align}
    S_{A}^{(n)}	
	&=\frac{1}{1-n}\Tr[\ln[(\I-\bm{C})^{n}+\bm{C}^{n}]],\mylabel{eq:Sn_nI} 
\end{align}
in agreement with the known results \cite{Casini2009}. Taking limit $n\to 1$ above, we get the \response{subsystem von Neumann entropy}, $S_{EE}=-\Tr[(\I-C)\log(\I-C)+C\log C]$.

As shown in \appcite{sec:Keldysh-nI}, in the non-equilibrium case, using the Schwinger-Keldysh path integral, discussed in \Seccite{para:Keldysh-chi}, we obtain the same expression for the $n$-the R\'{e}nyi entropy $S_{A}^{(n)}(t)$ as given in \Eqn{eq:Sn_nI} with a time-dependent correlation matrix
\begin{align}\mylabel{eqn:cor-mat-keldysh}
    C_{ij}(t)=\Tr[\rho(t)\cd_ic_j].
\end{align}
Here $\rho(t)$ is the density matrix at time $t$.

Having reproduced the results for non-interacting fermionic systems, we now seek to set up the field theory for R\'{e}nyi entropy in some well-known models for interacting fermions in the next sections. We discuss the formulation for two types of systems, large-$N$ fermionic models based on SYK model \cite{Sachdev1993,KitaevKITP} and repulsive Hubbard model treated within single-site DMFT \cite{Georges1996RMP}.

\subsection{R\'{e}nyi entropy for SYK and related models}\mylabel{sec:allmodels}
\begin{figure}
    \centering
    \includegraphics[width=0.5\textwidth]{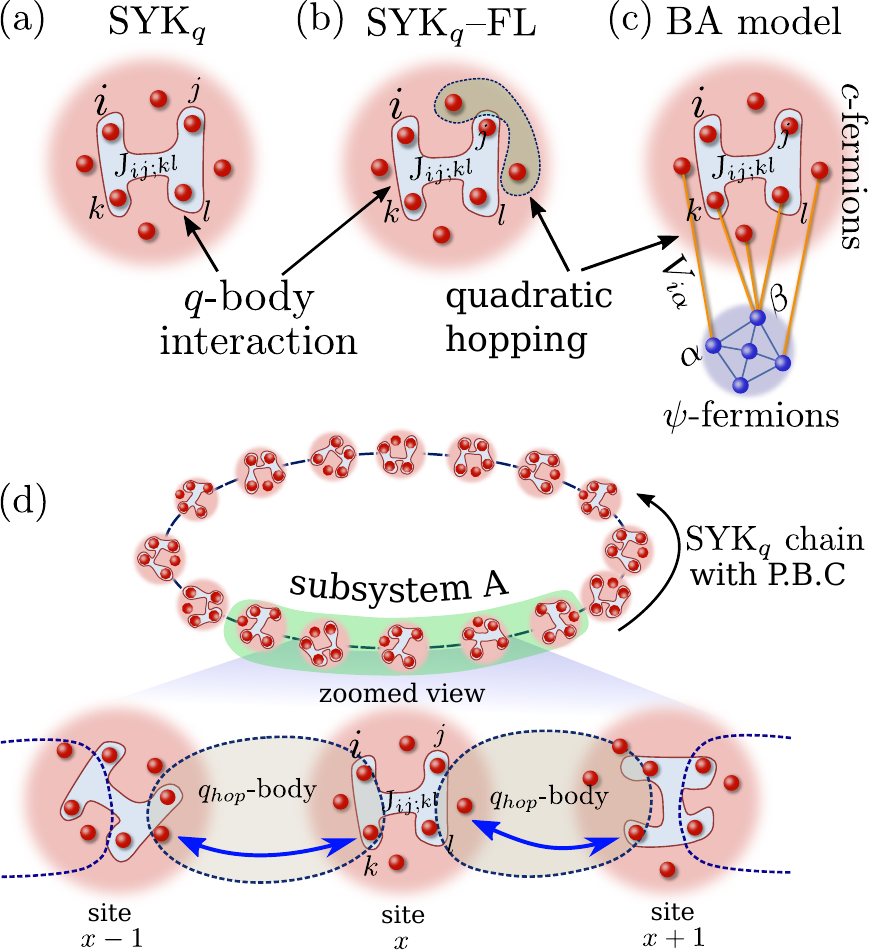}
    \caption{{\bf Large-$N$ models:} (a) The $q$-body SYK model (SYK$_q$) for complex-fermions shown for $q=2$ interactions, that scatter a pair of fermions, for e.g., from sites $k$, $l$ to sites $i$, $j$ with a random complex amplitude $J_{ijkl}$. (b) Model for a zero-dimensional interacting-Fermi liquid obtained from the SYK$_q$ model by adding quadratic (one-body) hopping terms (represented by dashed lines) on top of  the $q$-body ($q\geq 2)$ interactions. (c) The BA model for NFL-FL transition, obtained by joining a non-interacting dot of peripheral $\psi$-fermions (blue dots joined by lines) to the SYK-dot of $c$-fermions via random hopping amplitudes $V_{i\alpha}$. (d) A chain of SYK$_{q}$ dots with periodic boundary conditions (P.B.C). Each dot is connected to its nearest neighbors via $q_{hop}$-body terms, e.g. $q_{hop}=1$, as shown in the zoomed view (below). The highlighted part of the chain is chosen as the subsystem  $A$ for which entanglement-entropy is calculated.}
    \label{fig:SYK-models}
\end{figure}
SYK model \cite{Sachdev1993,KitaevKITP} and its various extensions \cite{Gu2017,BanerjeeAltman2016,SJian2017b,Song2017,Davison2017,Patel2018,Chowdhury2018,Arijit2017,Haldar2018} have  emerged in recent years as a new paradigm to study phases of strongly interacting fermions, e.g. NFL, marginal and heavy Fermi liquids, strong-coupling superconductivity \cite{Song2017,Patel2018,Chowdhury2018,Arijit2017,Haldar2018,Hauck2020,Chowdhury2020} and quantum phase transitions \cite{BanerjeeAltman2016,CJian2017,Haldar2018,Arijit2017,Can2019PRB}. Remarkably, these models can be solved exactly in a suitable large-$N$ limit, without resorting to any kind of perturbative treatment. As a result, the study of this family of large-$N$ fermionic model have enabled new insights into transport \cite{Gu2017,SJian2017b,Song2017,Davison2017,Patel2018,Chowdhury2018}, thermalization \cite{Eberlein2017,Sonner2017,Kourkoulou2017,Bhattacharya2018,Haldar2020,Zhang2019,Almheiri2019,Maldacena2019,Kuhlenkamp2020}, many-body chaos \cite{KitaevKITP,Kitaev2018,BanerjeeAltman2016} and entanglement \cite{Gu2017,LiuPRB2018,Huang2019,Chen2019,Zhang2020,Zhang2020a} in strongly interacting fermionic systems, as well as their intriguing connections with black holes \cite{Sachdev1993,KitaevKITP,Maldacena2016}.
Here we use the path integral technique of \Seccite{subsec:RenyiFieldTheory} to formulate the field theory for the $2$-nd R\'{e}nyi entropy for the SYK-model and its various generalizations discussed below. 

\emph{SYK model for non-Fermi liquid state:} The SYK model describes a zero-dimensional system having $N$ fermion flavors or sites (see \Fig{fig:SYK-models}(a)), that interact via an all-to-all or infinite-range Hamiltonian.
To keep the discussion general, we use the $q$-body version \mycite{Maldacena2016} of the SYK model, $\mathrm{SYK}_q$ (typically referred in the literature as $\mathrm{SYK}_{2q}$)  described by the Hamiltonian
\begin{align}\mylabel{eq:SYKqham}
    H_{SYK_q}=\sum_{i_{1}\cdots i_{q};j_{1}\cdots j_{q}}J_{i_{1}\cdots i_{q};j_{q}\cdots j_{1}}c_{i_{1}}^{\dagger}\cdots c_{i_{q}}^{\dagger}c_{j_{q}}\cdots c_{j_{1}},
\end{align}
where $J_{i_{1}\cdots i_{q};j_{q}\cdots j_{1}}$ are properly antisymmetrized Gaussian random numbers with variance $J^2/qN^{2q-1}(q!)^2$. Setting $q=2$ in the above Hamiltonian gives back the original SYK model \cite{Sachdev1993,KitaevKITP}. The ground state for the SYK$_q$ model, for $q\geq2$, has been shown to be a non-Fermi Liquid (NFL), lacking quasi-particle excitations, with a fermion scaling dimension $\Delta=1/2q$ \cite{Maldacena2016}. Interestingly, the SYK NFL states possess a residual zero-temperature thermodynamic entropy $S_0$ in the limit $N\to \infty$, e.g $S_0\approx 0.464$ for $q=2$, which, as we shall see later, plays a crucial role in the interpretation of the \response{$T=0$ R\'{e}nyi entropy in the large-$N$ limit} as well. The $q=1$ model is a special case whose ground state is a \emph{non-interacting} Fermi Liquid described by a single-particle semi-circular density of states (DOS). 

\emph{SYK model with quadratic hopping term for Fermi liquid state:} The SYK model has also been generalized to describe \emph{strongly} interacting or heavy Fermi-liquids (FL) \mycite{Song2017} by adding a random all-to-all hopping ($q=1$) term to \Eqn{eq:SYKqham}, i.e.,
\begin{align}\mylabel{eq:SYKFLham}
   H_{FL}=&\sum_{ij}t_{ij}\cd_i c_j\notag\\
   &+ \sum_{i_{1}\cdots i_{q};j_{1}\cdots j_{q}}J_{i_{1}\cdots i_{q};j_{q}\cdots j_{1}}c_{i_{1}}^{\dagger}\cdots c_{i_{q}}^{\dagger}c_{j_{q}}\cdots c_{j_{1}}
\end{align}
where $t_{ij}$ are the Gaussian random numbers with variance $\thop^2/N$ (see \Fig{fig:SYK-models}(b)). 

\emph{BA model for non Fermi liquid to Fermi liquid transition:} We also study  R\'{e}nyi entropy in the model of ref.\cite{BanerjeeAltman2016}, $\mc{H}=\mc{H}_c+\mc{H}_{\psi}+\mc{H}_{c\psi}$, where
\begin{subequations} \label{eq:BAModel}
\bea
\mc{H}_c&=&\sum_{ijkl}J_{ijkl}c_{i}^{\dagger}c_{j}^{\dagger}c_{k}c_{l}
\label{eq.SYK}
\\
\mc{H}_{\psi}&=&\sum_{\alpha\beta}t_{\alpha\beta}\psi_{\alpha}^{\dagger}\psi_{\beta}
\label{eq.Anderson}\\
\mc{H}_{c\psi}(t)&=&\sum_{i\alpha}(V_{i\alpha}c_{i}^{\dagger}\psi_{\alpha}+V_{i\alpha}^{*}\psi_{\alpha}^{\dagger}c_{i}).
\label{eq.Coupling}
\eea 
\end{subequations}
The above model (see \Fig{fig:SYK-models}(c)) has two species of fermions -- (1) the SYK fermions ($c$), on sites $i=1,\dots,N_c$, interacting with random coupling $J_{ijkl}$ (\Eqn{eq.SYK}) with variance $J^2/(2N_c)^{3/2}$; and (2) the \emph{peripheral} fermions ($\psi$), on a separate set of sites $\alpha=1,\dots,N_\psi$ connected via random all-to-all hopping $t_{\a\b}$ (\Eqn{eq.Anderson}). The SYK and the peripheral fermions  are quadratically coupled via $V_{i\a}$; $t_{\a\b}$ and $V_{i\a}$ are Gaussian random variables with variances $\thop^2/N_c$ and $V^2/\sqrt{N_cN_\psi}$, respectively. 

The model is exactly solvable for $N_c,N_\psi\to\infty$ with a fixed ratio $p=N_\psi/N_c$, that is varied to go through the QPT between NFL and FL at a critical value $p=p_c=1$ \cite{BanerjeeAltman2016}. The residual entropy density $S_0(p)$ of the SYK NFL continuously vanishes at the transition \cite{BanerjeeAltman2016}.

\emph{Lattice of SYK dots for interacting diffusive metal:} Extension of the above zero-dimensional models to higher dimensions have also been achieved \mycite{Gu2016,Davison2017,Song2017,Arijit2017,Chowdhury2018}. These systems typically involve a lattice of SYK$_q$ dots, each having $N$ fermion flavors, connected to their nearest neighbors via $q_{hop}$-body terms and can be described using the general Hamiltonian 
\begin{align}\mylabel{eq:HD-SYK-chain}
    H&=\sum_{\langle xx'\rangle i_{1}\cdots i_{q};j_{1}\cdots j_{q}}t_{x,i_{1}\cdots i_{q};x'j_{q}\cdots j_{1}}c_{xi_{1}}^{\dagger}\cdots c_{xi_{q}}^{\dagger}c_{x'j_{q}}\cdots c_{x'j_{1}}\notag\\
    &+h.c.\notag\\
    &+\sum_{x,i_{1}\cdots i_{q};j_{1}\cdots j_{q}}J_{x,i_{1}\cdots i_{q};j_{q}\cdots j_{1}}c_{xi_{1}}^{\dagger}\cdots c_{xi_{q}}^{\dagger}c_{xj_{q}}\cdots c_{xj_{1}},
\end{align}
where the coordinates $x$, $x'$ denote the position of the individual dot in the lattice or chain and $\langle x x'\rangle$ represents the nearest neighbors (see \Fig{fig:SYK-models}(d)). The amplitude $t_{xx'}$ and $J_x$ are independent Gaussian random numbers with variances having the same form as that of the coupling $J$ in \Eqn{eq:SYKqham}; $q\to q_{hop}$ and $J\to \tchain$ for the $t_{xx'}$ amplitudes. Setting $q_{hop}=q=1$ produces a non-interacting \emph{diffusive metal}, while $q_{hop}=1$, $q=2$ and $q_{hop}=2$, $q=2$ results in a diffusive heavy-Fermi liquid \cite{Song2017} or a non-Fermi liquid \cite{Gu2017}, respectively.

\subsubsection{R\'{e}nyi entropy in the thermal state}\mylabel{sec:largeNthermaltheory}
Here we derive the exact equations to evaluate the R\'{e}nyi entropy for all the models of the preceding section at the large-$N$ limit in thermal equilibrium. The evaluation of R\'{e}nyi entropy for non-equilibrium evolution can be performed using the Schwinger-Keldysh path integral formalism of \Seccite{para:Keldysh-chi} and is discussed in \appcite{app:keldysh-SYK}.
In order to access the second \response{R\'{e}nyi entropy at zero temperature} we use the thermal density matrix to perform the analysis and then take the $T\to 0$ limit.
The definition for $2$-nd R\'{e}nyi entropy (see \Eqn{eq:n-Renyi-def}) implies, that we need to calculate the disorder averaged quantity
\begin{align}\mylabel{eq:S2-def-thermal-SYK}
    S&^{(2)}=-\overline{\ln\left(\Tr_{A}[\rho_{A}^{2}]\right)} =-\overline{\ln\Tr_{A}[Z_{A}^{2}]}+2\overline{\ln Z},
\end{align}
where $\overline{\cdots}$ represents disorder average over the amplitudes $J_{i_{1}\cdots i_{q};j_{q}\cdots j_{1}}$. We also define the operator $Z_A=\Tr_B \exp(-\beta H)$.
We evaluate each term in \Eqn{eq:S2-def-thermal-SYK} separately. Using the trace identity (see \Eqn{eq:Tr-FG-exp-DN}) we formally write down $\Tr_A[Z_A^2]$ as
\begin{align}\mylabel{eq:TrZA2-expansion}
\Tr_{A}[Z_{A}^{2}]=\int d^{2}(\bs{\xi},\bs{\eta})\ f_{N}(\bs{\xi},\bs{\eta})\mathrm{Tr}[ZD_{N}(\bs{\xi})]\mathrm{Tr}[ZD_{N}(\bs{\eta})],    \end{align}
where as before $\bs{\xi}$ and $\bs{\eta}$ couple only to sites in the region $A$, and $\Tr[ZD_{N}(\bs{\xi})]$ is the characteristic function for the partition function $Z$. Since we need to evaluate the logarithm of $\Tr_A[Z_A^2]$ (see \Eqn{eq:S2-def-thermal-SYK}), we use the \emph{replica}-trick to express the log as a product of \emph{disorder replicas}
\begin{align}
    \overline{\ln\Tr_{A}[Z_{A}^{2}]}=\lim_{r\to0}\frac{\overline{\Tr_{A}[Z_{A}^{2}]^{r}}-1}{r}.
\end{align}
The replicas, along with \Eqn{eq:TrZA2-expansion} and the integral representation in \Eqn{eq:chi-path-int}, allows us to write the path-integral for the SYK$_q$ model (\Eqn{eq:SYKqham})
\begin{widetext}
\begin{align}\mylabel{eq:SYK-ZAr}
    &\Tr_{A}[Z_{A}^{2}]^{r}=2^{N_{A}r}\int \mathcal{D}(\cb,c)d^2(\bs{\xi},\bs{\eta})
    \exp\left[-\int_{0}^{\beta}\dtau\sum_{i,\sigma,a=1}^{r}\cb_{i\sigma a}\deltau c_{i\sigma a}\right.
    +\sum_{i\in A,\sigma,a}\int_{0}^{\beta}\dtau\left[ \delta(\tau^{+})\bar{c}_{i\sigma a}(\tau)\xi_{i\sigma a}-\bar{\xi}_{i\sigma a} \delta(\tau)c_{i\sigma a}(\tau)\right]\notag\\
    &\left.-\int_{0}^{\beta}\dtau\sum_{i,j,\sigma,a}
    J_{i_{1}\cdots i_{q};j_{q}\cdots j_{1}}
    \cb_{i_{1}\sigma a}\cdots\cb_{i_{q}\sigma a}
    c_{j_{q}\sigma a}\cdots c_{j_{1}\sigma a}\right]
    \exp\left[\frac{1}{2}\sum_{i\in A,a}(\bar{\xi}_{ia}\eta_{ia}-\bar{\eta}_{ia}\xi_{ia}-\bar{\xi}_{ia}\xi_{ia}-\bar{\eta}_{ia}\eta_{ia})\right].
\end{align}
\end{widetext}
The Grassmann fields $\cb_{i\sigma a}(\tau),c_{i\sigma a}(\tau)$, appearing in the path-integral above and representing the fermions in the model, are labeled by several indices apart from the site indices $i,j=1,\cdots, N$ -- an imaginary-time coordinate $\tau$, a disorder replica index $a=1,\cdots r$ and an \emph{entanglement replica} index $\sigma=1,2$. Here  $\xi_{i1a}=\xi_{ia}$ and $\xi_{i2a}=\eta_{ia}$, indicating the Grassmann-field originates either from the characteristic function $\Tr[ZD(\bs{\xi})]$ or $\Tr[ZD(\bs{\eta})]$, respectively in \Eqn{eq:TrZA2-expansion}. The Grassmann-variables $\xi$ and $\eta$ have the same indices as $\cb$,$c$ except for the time-coordinate $\tau$. The subsystem or the region $A$ in the zero-dimensional model is defined as any of the $N_A$ sites, e.g. $i=1,\dots,N_A$, out of the total $N$ sites.
Having obtained the path-integral, we now seek to evaluate the integral at the large-$N$ saddle point, i.e. $N\to\infty$. This limit can be accessed by taking the disorder average over $J_{i_{1}\cdots i_{q};j_{q}\cdots j_{1}}$s \rem{(see \appcite{})} and then introducing the large-$N$ field 
\begin{align}
    G_{\sigma'b,\sigma a}(\tau_{2},\tau_{1})=\frac{1}{N}\sum_{i}\cb_{i\sigma a}(\tau_{1})c_{i\sigma'b}(\tau_{2}),
\end{align}
and self-energy $\Sigma_{\sigma'b,\sigma a}(\tau_{2},\tau_{1})$ for the fermions. We point out that the above analysis closely follows that of the thermal-field theory of the SYK model. The latter can be found in refs.\mycite{Gu2017,Sachdev2015,BanerjeeAltman2016,Arijit2017,Haldar2018}, to name a few. The end result of introducing large-$N$ fields is an integral of the form
\begin{align}
    \overline{\Tr[Z_A^2]^r}=\int \mathcal{D}(\cb,c,\Sigma,G)d^2(\bs{\xi},\bs{\eta})e^{-\mathcal{S}[\cb,c,\bs{\xi},\bs{\eta},\Sigma,G]},
\end{align}
where the effective action $\mathcal{S}$ for the second R\'{e}nyi Entropy is bilinear in the Grassmann-fields $\cb,c$, $\bs{\xi}$, and $\bs{\eta}$, i.e.
\begin{align}\mylabel{eq:EE-action}
\mathcal{S}=&\left(-\sum_{i\in A,a=1}^{r}\left[\begin{array}{cc}
\bar{\xi}_{ia} & \bar{\eta}_{ia}\end{array}\right]\left[\begin{array}{cc}
\frac{1}{2} & -\frac{1}{2}\\
\frac{1}{2} & \frac{1}{2}
\end{array}\right]\left[\begin{array}{c}
\xi_{ia}\\
\eta_{ia}
\end{array}\right]\right.\nonumber\\
&+\sum_{i\in A,a=1}^{r}\int_{0}^{\beta}d\tau \delta(\tau^{+})\left[\begin{array}{cc}
\bar{c}_{i1a}(\tau) & \bar{c}_{i2a}(\tau)\end{array}\right]\left[\begin{array}{c}
\xi_{ia}\\
\eta_{ia}
\end{array}\right]\nonumber \\
&-\sum_{i\in A,a=1}^{r}\left.\int_{0}^{\beta}d\tau \delta(\tau)\left[\begin{array}{cc}
\bar{\xi}_{ia} & \bar{\eta}_{ia}\end{array}\right]\left[\begin{array}{c}
c_{i1a}(\tau)\\
c_{i2a}(\tau)
\end{array}\right]\right)\notag\\
&+\cdots.
\end{align}
The Gaussian structure, allows us to integrate the $\bs{\xi}$ and $\bs{\eta}$ variables, changing the effective action to another bilinear of $\cb$ and $c$
\begin{align}\mylabel{eq:EE-action-kick}
    \frac{1}{2^{N_{A}r}}\exp\left(-\int\sum_{i\in A,a}\bm{\bar{c}}_{ia}^{T}(\tau_{1})\underbrace{\begin{bmatrix}\begin{array}{rr}
1 & 1\\
-1 & 1
\end{array}\end{bmatrix}\delta(\tau_{1}^{+})\delta(\tau_{2})}_{\bm{M}_{\sigma_{1}\sigma_{2}}(\tau_{1},\tau_{2})}\bm{c_{ia}}(\tau_{2})\right),
\end{align}
where the matrix $M$ is defined as
\begin{align}\mylabel{eq:M-def}
    \bm{M}_{\sigma_{1}\sigma_{2}}(\tau_{1},\tau_{2})=\begin{bmatrix}\begin{array}{rr}
1 & 1\\
-1 & 1
\end{array}\end{bmatrix}\delta(\tau_{1}^{+})\delta(\tau_{2}),
\end{align}
and $c_{ia}(\tau)=\left[c_{i1a}(\tau)\ c_{i2a}(\tau)\right]^T$.
Interestingly, at this point we see explicitly, from \Eqn{eq:EE-action-kick}, how the fields $\bs{\xi}$, $\bs{\eta}$ extract the information regarding entanglement. Instead, of introducing complicated imaginary-time boundary conditions for fermions in sub-region $A$, the fields alter the self-energy by providing a time-dependent ``kick" to the  fermions, while leaving the fermions outside $A$ untouched. All the fermions can now be integrated to produce $\overline{\Tr[Z_A^2]^r}=\int \mathcal{D}(\cb,c,\Sigma,G)e^{-N\mathcal{S}[\Sigma,G]}$ and the final effective action
\begin{subequations}\mylabel{eq:action-before-freeE}
\begin{align}
    \mathcal{S}=&-(1-p)\ln\det\left(\bs{\deltau}+\bs{\Sigma}\right)
    -p\ln\det\left(\bs{\deltau}+\bs{\Sigma}+M\right)\notag\\
    &+\mathcal{S}_G, \\
	\mathcal{S}_G=&-\sum_{ab}^{r}\int\dtau_{1,2}(-1)^{q}\frac{J^{2}}{2q}G^{q}_{\sigma'b,\sigma a}(\tau_{2},\tau_{1})G^{q}_{\sigma a,\sigma' b}(\tau_{1},\tau_{2})\notag\\
	&-\sum_{ab}^{r}\int\dtau_{1,2}\Sigma_{\sigma a,\sigma' b}(\tau_{1},\tau_{2})G_{\sigma'a,\sigma b}(\tau_{2},\tau_{1}),
\end{align}
\end{subequations}
which depends only on the Green's function $G$ and self-energy $\Sigma$. Here $\int d\tau_{1,2}\equiv \int_0^\beta d\tau_1d\tau_2$. The symbol $p$ denotes the the ratio of subsystem size to total size, i.e.
\begin{align}
p=N_{A}/N \mylabel{eq:p-def-SYK-S2},   
\end{align}
 and can take a value from $0$ to $1$. The symbols $\bs{\partial_{\tau}}$, $\bs{\Sigma}$, appearing above, represents matrices having elements $\partial_{\tau_{1}}\delta(\tau_{1}-\tau_{2})\delta_{\sigma\sigma'}\delta_{ab}$, $\Sigma_{\sigma a,\sigma'b}(\tau_{1},\tau_{2})$ respectively. We evaluate the action $\mathcal{S}$ at the (disorder) replica diagonal and replica symmetric saddle point (i.e. $\Sigma,G\propto \delta_{ab}$) by minimizing with respect to $G$ and $\Sigma$ to get
\begin{align}\mylabel{eq:SP-SYK}
\bs{G}	=&(1-p)\bs{\tilde{G}}+p\bs{g}\notag\\
\bs{\tilde{G}}	=&-(\bs{\deltau}+\bs{\Sigma})^{-1}\notag\\
\bs{g}	=&-(\bs{\deltau}+\bs{\bs{\Sigma}}+\bs{M})^{-1}\notag\\
\Sigma_{\sigma\sigma'}(\tau_{1},\tau_{2})	=&(-1)^{q+1}J^{2}G_{\sigma\sigma'}(\tau_{1},\tau_{2})^{q}G_{\sigma'\sigma}(\tau_{2},\tau_{1})^{q-1},    
\end{align}
as the saddle point conditions. Here $\bs{G}$ is the matrix representation of the Green's function $G$ and $\bs{g}$, $\bs{\tilde{G}}$ are additional matrices that we have introduced to simplify the notation. Due to the similarities of the entanglement action $\mathcal{S}$ in \Eqn{eq:action-before-freeE} with the thermal free energy, we define the \emph{entanglement free-energy}
\begin{align}
    &F_{EE}(p,\beta)	=\frac{1}{2\beta}\mathcal{S}\notag\\
    &=\frac{1}{2\beta}\left[p\ln\det\left(-\mathbf{g}\right)+(1-p)\ln\det\left(-\mathbf{\tilde{G}}\right)\right.\notag\\
	&-\int_{0}^{\beta}\dtau_{1,2}\sum_{\sigma=1,2}(-1)^{q}\frac{J^{2}}{2q}G_{\sigma'\sigma}(\tau_{2},\tau_{1})^{q}G_{\sigma\sigma'}(\tau_{1},\tau_{2})^{q}\notag\\
	&\left.-\int_{0}^{\beta}\dtau_{1,2}\sum_{\sigma=1,2}\Sigma_{\sigma\sigma'}(\tau_{1},\tau_{2})G_{\sigma'\sigma}(\tau_{2},\tau_{1})\right].
\end{align}
which allows us to express the $2$-nd R\'{e}nyi entropy (\Eqn{eq:S2-def-thermal-SYK}) as 
\begin{align}\mylabel{eq:S2pbetadef}
    S^{(2)}(p,\beta)=2\beta\left(F_{EE}(p,\beta)-F_{EE}(p=0,\beta)\right),
\end{align}
which depends on $p$ and temperature $\beta^{-1}$. Note that we have used the fact that thermal free energy $F(\beta)$ at the saddle-point is given by $-\beta^{-1}\overline{\ln Z}$ and is equal to $F_{EE}(p=0,\beta)$ in \Eqn{eq:S2pbetadef}. The ground state R\'{e}nyi entanglement entropy for an arbitrary subsystem size $p$ can now be calculated by taking the limit
\begin{align}
    S^{(2)}(p)=\lim_{\beta\to\infty}S^{(2)}(p,\beta).
\end{align}
The above analysis can also be carried out for the generalizations of the SYK model given in Eqns.\eqref{eq:SYKFLham},\eqref{eq:HD-SYK-chain} and \eqref{eq:BAModel}. The equations to evaluate the R\'{e}nyi entropy for the strongly-interacting Fermi liquid Hamiltonian in \Eqn{eq:SYKFLham} and BA Hamiltonian in \Eqn{eq:BAModel} can be similarly obtained and are given in \appcite{app:OtherModels}. 

For extended systems of the kind described by \Eqn{eq:HD-SYK-chain}, i.e. a ring of SYK dots connected to their nearest neighbors, we choose the subsystem as the $l$ successive dots, e.g. $x=1,\dots,l$ (see \Fig{fig:SYK-models}(d)) as the subsystem $A$. The saddle-point conditions for this arrangement are 
\begin{align}\mylabel{eq:large-N-1D-SP}
    (\deltau+\bs{\Sigma}^{(x)}&+M\delta_{x\in A})\bs{G}^{(x)}	=-1\notag\\
\Sigma^{(x)}_{\sigma\sigma'}(\tau_{1},\tau_{2})	=&(-1)^{q+1}J^{2}G^{(x)}_{\sigma\sigma'}(\tau_{1},\tau_{2})^{q}G^{(x)}_{\sigma\sigma'}(\tau_{2},\tau_{1})^{q-1}\notag\\
&(-1)^{q_{hop}+1}\tchain^{2}\left[G^{(x-1)}_{\sigma\sigma'}(\tau_{1},\tau_{2})^{q_{hop}}\right.\notag\\
&+\left.G^{(x+1)}_{\sigma\sigma'}(\tau_{1},\tau_{2})^{q_{hop}}\right]G^{(x)}_{\sigma'\sigma}(\tau_{2},\tau_{1})^{q_{hop}-1},
\end{align}
where the matrix $M$ (defined in \Eqn{eq:M-def}) adds to the site dependent self-energy -- $\Sigma^{(x)}$ only for dots which belong to subsystem $A$, as indicated by the Kronecker delta function $\delta_{x\in A}$. The Green's function $G^{(x)}$ also become dependent on the site index $x$. The entanglement free energy can be calculated in terms of the space dependent $G^{(x)}$ and $\Sigma^{(x)}$, i.e.
\begin{align}\mylabel{eq:FEE-chain}
    &F_{EE}(p,\beta)\notag\\	
    &=\frac{1}{2\beta}\left[\sum_{x}\ln\det\left(-\mathbf{G}^{(x)}\right)\right.\notag\\
	&-\int_{0}^{\beta}\dtau_{1,2}\left\{\sum_{x,\sigma}(-1)^{q}\frac{J^{2}}{2q}G_{\sigma'\sigma}^{(x)}(\tau_{2},\tau_{1})^{q}G_{\sigma\sigma'}^{(x)}(\tau_{1},\tau_{2})^{q}\right.\notag\\
	&+\sum_{x,\sigma}\Sigma_{\sigma\sigma'}^{(x)}(\tau_{1},\tau_{2})G_{\sigma'\sigma}^{(x)}(\tau_{2},\tau_{1})\notag\\
	&\left.\left.+\sum_{\langle xx'\rangle\sigma\sigma'}\frac{(-1)^{q_{hop}}\tchain^{2}}{q_{hop}}G_{\sigma\sigma'}^{(x)}(\tau_{1},\tau_{2})^{q_{hop}}G_{\sigma'\sigma}^{(x')}(\tau_{2},\tau_{1})^{q_{hop}}\right\}\right],
\end{align}
where $\langle x x'\rangle$ represents nearest neighbors. The presence of $\delta_{x\in A}$ in \Eqn{eq:large-N-1D-SP} explicitly breaks translation symmetry and one needs to retain the Green's functions for each site $x$ in order to calculate the entanglement entropy. 

We obtain the R\'{e}nyi entropy in \Seccite{sec:Results} for all the models discussed in this section by numerically solving the saddle-point equations, such as \Eqn{eq:SP-SYK} and \Eqn{eq:large-N-1D-SP}, and using \Eqn{eq:S2pbetadef}. We discuss the iterative numerical algorithm to solve the saddle-point equations, in \appcite{sec:sp-num}.

\rem{\textcolor{blue}{Analysis starts from where we left of in \Seccite{sec:formulation} and ends with Kadanoff-Baym equations and Keldysh action}}

\subsection{R\'{e}nyi entropy in Hubbard model: Dynamical mean-field theory (DMFT)} \label{subsec:DMFT}
In this section we demonstrate the application of the path integral method of \Seccite{subsec:RenyiFieldTheory} for the Hubbard model. We develop the formulation to compute R\'{e}nyi entropy within the dynamical mean field theory (DMFT) \cite{Georges1996RMP}, one of the successful approaches to treat electronic correlation and metal-insulator transition in Hubbard model. We formulate the DMFT for R\'{e}nyi entropy in the single-site approximation. In the usual single-site DMFT \cite{Georges1996RMP} one reduces a lattice problem into an impurity coupled to a bath and the properties of the impurity and the bath are calculated self-consistently using the non-interacting dispersion. We extend the DMFT approximation to evaluate the path integrals developed in \Seccite{subsec:RenyiFieldTheory}. The method can be integrated with the continuous-time quantum Monte Carlo (CTQMC) impurity solver \cite{Gull2011} and extended to the cluster implementations \cite{Maier2005}.  Here we only discuss the formulation of the problem within DMFT, the numerical implementation will be discussed in a future publication \cite{Banerjee2020}.

We consider the Hubbard model
\begin{align}
H= & \sum_{(ij),\sigma}t_{ij}c_{i\sigma}^{\dagger}c_{j\sigma}-\mu\sum_i n_i+U\sum_{i}n_{i\uparrow}n_{i\downarrow} \label{eq:HubbardModel}
\end{align}
with hopping amplitudes $t_{ij}$ between sites on a lattice, $\mu$ the chemical potential and $U$ the on-site repulsive interaction between fermions with opposite spins $\sigma=\uparrow,\downarrow$. Here $n_{i\sigma}=c_{i\sigma}^\dagger c_{i\sigma}$ and $n_i=\sum_\sigma n_{i\sigma}$ are the electronic number operators.

We discuss the second R\'{e}nyi entropy of a subsystem for equilibrium state described by the thermal density matrix (\Eqn{eq:rho-therm}). The method can be easily extended to non-equilibrium situation (\Seccite{subsec:RenyiFieldTheory}) via non-equilibrium DMFT \cite{Aoki2014}. As in \Eqn{eq:S2-def-thermal-SYK}, we compute
\begin{align} \label{eq:GrandPot}
S^{(2)}&=\beta(\Omega^{(2)}-2\Omega),
\end{align}
where $\Omega^{(2)}=-T\ln Z^{(2)}$ ($Z^{(2)}\equiv\Tr_AZ_A^2$) and $\Omega=-T\ln Z$ are relevant grand potentials. Again, using the trace identity (\Eqn{eq:TrZA2-expansion}) we obtain the path integral, 
\begin{subequations}
\label{eq:HubbardAction}
\begin{align} 
&Z^{(2)}= \int\prod_{i\in A,\sigma,\alpha=1,2}d^{2}\xi_{i\sigma\alpha}f_N(\bs{\xi})\int\mathcal{D}(\bar{c},c)e^{-\mathcal{S}}\label{eq:Z2-1}\\
&\mathcal{S}= \int_{0}^{\beta}d\tau\left[\sum_{i\sigma\alpha}\bar{c}_{i\sigma\alpha}\left[(\partial_{\tau}-\mu)\delta_{ij}+t_{ij}\right]c_{j\sigma\alpha}\right.\nonumber\\
&+\left. U\sum_{i\alpha}n_{i\uparrow\alpha}n_{i\downarrow\alpha}\right]\nonumber\\
&+\sum_{i\in A,\sigma\alpha}\int_{0}^{\beta}d\tau\left[\bar{\xi}_{i\sigma\alpha} \delta(\tau)c_{i\sigma\alpha}(\tau)-\delta(\tau^+)\bar{c}_{i\sigma\alpha}(\tau)\xi_{i\sigma\alpha}\right] 
\end{align}
\end{subequations}
Here we have the auxiliary Grassmann fields $\bs{\xi}=\{\bar{\xi}_{i\sigma\alpha},\xi_{i\sigma\alpha}\}_{i\in A}$ with spin ($\sigma$) and entanglement replica ($\alpha=1,2$) indices, and $f_N(\bs{\xi})=\exp[-(1/2)\sum_{i\in A,\sigma}(\bar{\xi}_{i\sigma 1}\xi_{i\sigma 1}+\bar{\xi}_{i\sigma 2}\xi_{i\sigma 2}-\bar{\xi}_{i\sigma 1}\xi_{i\sigma 2}+\bar{\xi}_{i\sigma 2}\xi_{i\sigma 1})]$. As in the case of chain of SYK dots in \Seccite{sec:largeNthermaltheory}, due to the entanglement subdivision, the above action breaks translational invariance. Hence, we use a single-site inhomogeneous DMFT to reduce the lattice problem (\Eqn{eq:HubbardModel}) into an impurity problem for site $i$,
\begin{align}\label{eq:ImpurityModel}
&Z_{i}^{(2)}= \int\left(d^{2}\bs{\xi} f_N(\bs{\xi})\right)^{\delta_{i\in A}}\mathcal{D}(\bar{c},c)e^{-\mathcal{S}_{i}}\\
&\mathcal{S}_{i}=-\int d\tau_{1,2}\sum_{\sigma\alpha\gamma}\bar{c}_{\sigma\alpha}(\tau_1)\mathcal{G}_{i\alpha\gamma}^{-1}(\tau_1,\tau_2)c_{\sigma\gamma}(\tau_2)\nonumber\\
&+U\int_{0}^{\beta}d\tau\sum_{\alpha}n_{\uparrow\alpha}(\tau)n_{\downarrow\alpha}(\tau)\nonumber\\
 & +\delta_{i\in A}\sum_{\sigma\alpha}\int_{0}^{\beta}d\tau \left[\bar{\xi}_{\sigma\alpha}\delta(\tau)c_{\sigma\alpha}(\tau)-\delta(\tau^+)\bar{c}_{\sigma\alpha}(\tau)\xi_{\sigma\alpha}\right]
\end{align}
The auxiliary Grassmann fields $\{\bar{\xi}_{\sigma\alpha},\xi_{\sigma\alpha}\}$ only appears for sites belonging to the $A$ region; $d^2\bs{\xi}=\prod_{\sigma\alpha}d^2\xi_{\sigma\alpha}$ and $f_N(\bs{\xi})=\exp[-(1/2)\sum_{\sigma}(\bar{\xi}_{\sigma 1}\xi_{\sigma 1}+\bar{\xi}_{\sigma 2}\xi_{\sigma 2}-\bar{\xi}_{\sigma 1}\xi_{\sigma 2}+\bar{\xi}_{\sigma 2}\xi_{\sigma 1})]$. The above impurity action could be derived from \Eqn{eq:HubbardAction} using the cavity method \cite{Georges1996RMP}. To simplify the notations, we have assumed a paramagnetic state with spin symmetry, and hence the dynamical `Weiss mean field' is independent of spin and is given by
\begin{align}
\mathcal{\mathcal{G}}_{i}^{-1}(\tau,\tau')= & -(\partial_{\tau}-\mu)\delta(\tau-\tau')\mathcal{I}-\Delta_{i}(\tau,\tau'),
\end{align}
A $2\times 2$ matrix in the replica space, where $\mathcal{I}$ denotes a ($2\times2$) unit matrix and $\Delta_{i}(\tau,\tau')$ is the hybridization 
function which carries the information about the lattice and needs to be found self-consistently (see below). For $i\in A$, we integrate out the entangling Grassmann fields $\bar{\xi}_{\sigma\alpha},\xi_{\sigma\alpha}$ to get $Z_i^{(2)}=\int \mathcal{D}(\bar{c},c)e^{-\tilde{\mathcal{S}_i}}$ and the effective action
\begin{align}
\tilde{\mathcal{S}}_{i}&= -\int d\tau_{1,2}\sum_{\sigma\alpha\gamma}\bar{c}_{\sigma\alpha}(\tau_1)\tilde{\mathcal{G}}_{i\alpha\gamma}^{-1}(\tau_1,\tau_2)c_{\sigma\gamma}(\tau_2)\nonumber \\
&+U\int_{0}^{\beta}d\tau\sum_{\alpha}n_{\uparrow\alpha}(\tau)n_{\downarrow\alpha}(\tau)\\
\tilde{\mathcal{G}}_{i}^{-1}(\tau,\tau')&=  -(\partial_{\tau}-\mu)\delta(\tau-\tau')\mathcal{I}-\Delta_{i}(\tau,\tau')\nonumber \\
&-\delta_{i\in A}M(\tau,\tau'),
\end{align}
where the $2\times 2$ matrix $M$ is given in \Eqn{eq:M-def}. The impurity Green's function $G_i(\tau,\tau')$ can be obtained from the impurity Dyson equation,
\begin{align}
G^{-1}_i(\tau,\tau')=\tilde{\mathcal{G}}_{i}^{-1}(\tau,\tau')-\Sigma_i(\tau,\tau'),
\end{align}
where $\Sigma_{i,\alpha\gamma}(\tau,\tau')$ is the impurity self-energy that needs to be computed using an appropriate impurity solver, e.g. iterative perturbation theory (IPT) \cite{Georges1996RMP} or CTQMC \cite{Gull2011}. 
The hybridization function can be obtained in terms of cavity Green's function as
\begin{align}\label{eq:HybridFn}
\Delta_{i\alpha\gamma}(\tau,\tau')= & \sum_{jk}t_{ij}t_{ik}G_{j\alpha,k\gamma}^{(i)}(\tau,\tau').
\end{align}
Here $G^{(i)}(\tau,\tau')$ is the Green's function with $i$-th
site removed from the lattice. The cavity Green's function can be obtained from the lattice Green's functions as
\begin{align}
G_{j\alpha,k\gamma}^{(i)}(\tau,\tau')= & G_{j\alpha,k\gamma}(\tau,\tau')-\sum_{\eta\delta}\int d\tau_{1,2}\left[G_{j\alpha,i\eta}(\tau,\tau_{1})\right.\nonumber \\
&\left. G_{i\eta,i\delta}^{-1}(\tau_{1},\tau_{2})G_{i\delta,k\gamma}(\tau_{2},\tau')\right]\label{eq:CavityGnFn}
\end{align}
Where $G(\tau,\tau')$ is the full lattice Green's function. To obtain the full lattice Green's function, within single-site DMFT approximation, one assumes that the lattice self-energy is local and is given by the impurity self-energy. Then, the lattice Green's function is obtained from the Dyson equation
\begin{align}
&\sum_j\int d\tau_{1}\left[-\left\{(\partial_{\tau}-\mu)\delta_{ij}+t_{ij}\right\}\delta(\tau-\tau_{1})\mathcal{I}\right.\nonumber\\
&\left.+\delta_{i\in A}\delta_{ij}M(\tau,\tau_1)-\delta_{ij}\Sigma_{i}(\tau,\tau_{1})\right] G_{jk}(\tau,\tau')
\nonumber \\
&=\delta_{ik}\delta(\tau-\tau')\mathcal{I}\label{eq:latticeDyson}
\end{align}
The above equation can be used to obtain the hybridization function in \Eqn{eq:HybridFn} and thus closes the DMFT self-consistency loop.
The numerical solution of the above DMFT equations is rather involved since both space and time translation invariance are broken, the former due to the choice of subregion $A$ and the latter due to the time-dependent self-energy kick used to extract the entanglement. The numerical solution will be discussed in a future work \cite{Banerjee2020}. 

We conclude this section by briefly mentioning the important step to compute the grand potentials $\Omega^{(2)}$ and $\Omega$ appearing in Eq.\eqref{eq:GrandPot} for the second R\'{e}nyi entropy. The grand potentials can be obtained via coupling constant integration discussed in detail in \appcite{app:CouplingConstInt}. In this method, one considers a modified Hamiltonian $H_\lambda=H_0+\lambda H_1$, where $H_0$ is the non-interacting part in Eq.\eqref{eq:HubbardModel} and $H_1$ is the Hubbard term with the coupling constant $0\leq \lambda \leq 1$; $\lambda=0$ is the non-interacting limit and the $\lambda=1$ is the interacting Hamiltonian of interest. The grand potentials could be obtained as
\begin{align}
\Omega^{(2)}&= \Omega^{(2)}(0)\nonumber \\
&+\int_{0}^{1}\frac{d\lambda}{\lambda}\sum_{ij\sigma}\left[(\partial_{\tau_{0}}-\mu)\delta_{ij}+t_{ij}\right]G_{j\sigma\alpha,i\sigma\alpha}^{(\lambda)}(0,0^{+})\label{eq:GrandPotRenyi}\\
\Omega&= \Omega(0)\nonumber \\
&+\int_{0}^{1}\frac{d\lambda}{\lambda}\sum_{ij\sigma}\left[(\partial_{\tau}-\mu+\epsilon_{i})\delta_{ij}+t_{ij}\right]G_{j\sigma,i\sigma}^{(\lambda)}(\tau,\tau^{+})\label{eq:GrandPotThermo}
\end{align}
The integrands above can be calculated by computing the Green's function $G^{(\lambda)}$ via DMFT for each $\lambda$. One important point here is that the equal-time Green's function appearing in the expression for $\Omega^{(2)}$ needs to be calculated at the imaginary-time instant where the fermionic source fields are inserted in the path integral, in our case $\tau=0$. In contrast, $\tau$ in the expression of usual thermodynamic grand potential $\Omega$ is arbitrary. $\Omega^{(2)}(0)$ and $\Omega(0)$ are the grand potentials for the non-interacting system ($\lambda=0$) which can be calculated directly using the results of \Seccite{subsec:nI-formalism}.

\section{Analytical and numerical results for R\'{e}nyi entropy in large-$N$ fermionic models of Fermi liquids and non-Fermi liquids}\mylabel{sec:Results}
With the formalism for calculating \response{R\'{e}nyi} entropy for large-$N$ systems in place, we proceed forward and discuss the solutions obtained for the models introduced in \Seccite{sec:allmodels}. In this paper, we only describe the results for the $2$-nd R\'{e}nyi entropy obtained from the thermal field theory in \Seccite{sec:largeNthermaltheory}. The results for higher-order R\'{e}nyi entropies and time-evolution of R\'{e}nyi entropy in non-equilibrium situations will be communicated in a future work \cite{Haldar2020Renyi}. 
\subsection{Non-interacting large-$N$ model with disorder}
To set the stage and bench mark our large-$N$ field theory formalism of \Seccite{sec:largeNthermaltheory} we first discuss the R\'{e}nyi entropy in a non-interacting model. For $q=1$, the model defined in \Eqn{eq:SYKqham} describes a non-interacting system connected with disordered random hoppings having a variance $\thop^2$, and hence is amenable to solution in multiple ways. Therefore, we compare the temperature dependent R\'{e}nyi entropy, defined in \Eqn{eq:S2pbetadef}, calculated using -- (a) the correlation matrix approach applicable for non-interacting system as discussed in \Seccite{subsec:nI-formalism}, where the correlation matrix is evaluated by diagonalizing the non-interacting Hamiltonian, (b) using many-body exact diagonalization (ED) (see \appcite{sec:finite-N-num}) and (c) using the large-$N$ thermal field developed in \Seccite{sec:largeNthermaltheory}. For cases (a) and (b), the R\'{e}nyi entropy is calculated by explicitly averaging over multiple disorder realizations. 
\begin{figure}
    \centering
    \includegraphics[width=0.48\textwidth]{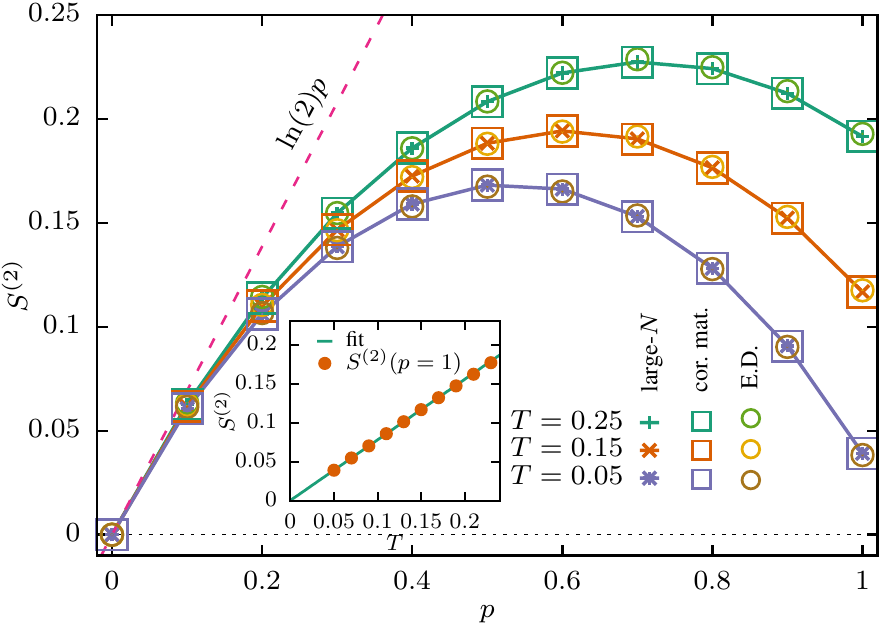}
    \caption{{\bf R\'{e}nyi entropy in non-interacting large-$N$ model:} Second R\'{e}nyi-Entropy, $S^{(2)}$, for the non-interacting model with random hoppings shown as a function of subsystem fraction $p$ for several values of temperature $T$. The predictions obtained from large-$N$ saddle-point theory (points with lines) show an excellent agreement with results (represented by color matched squares) obtained using the correlation matrix, as well as exact diagonalisation (ED) calculations (circles) performed for $N=10$ sites. All three techniques show a maximal entanglement scaling of R\'{e}nyi entropy for small subsystem sizes ($p\to 0$) with a $\ln(2)$ coefficient. Inset, gives the temperature dependence of the R\'{e}nyi entropy at $p=1$, i.e. $S^{(2)}(p=1)$, which goes to \emph{zero} as temperature is decreased indicating that the thermal density matrix, used for calculating entanglement, approaches the ground state for the model.}
    \label{fig:nI-large-N}
\end{figure}

\Fig{fig:nI-large-N} shows the variation of $2$-nd R\'{e}nyi entropy with subsystem size $p$ for multiple temperatures calculated using the above methods. We get an excellent match for all three cases. Particularly remarkable is the fact that the large-$N$ results almost exactly match with that obtained with ED for quite small system size ($N=10$), implying that $1/N$ corrections to entanglement entropy is rather small. This feature persists for the interacting large-$N$ models discussed in the next sections. We find that for small subsystem size, i.e. $p\to 0$, the R\'{e}nyi entropy grows with $p$ with a slope of $\ln(2)$ (see dashed line in \Fig{fig:nI-large-N}) indicating a maximal entanglement which can be argued based on the counting of degrees of freedom \mycite{Fu2016,LiuPRB2018}. In the limit of large subsystem size, i.e. $p\to 1$, the \response{R\'{e}nyi} entropy attains a finite \emph{non-zero} value that goes to \emph{zero} as $T\to 0$, see \Fig{fig:nI-large-N} inset. At finite temperature the R\'{e}nyi entropy of the subsystem for a thermal mixed-state density matrix picks up contribution from usual thermal entropy, which is only of statistical origin. However, as temperature is lowered ($\beta\to \infty$) the thermal density matrix approaches that of the ground-state, i.e. $\rho^2\to\rho$.
For any pure-state, in this case the ground state, the entanglement entropy goes to \emph{zero} as the subsystem size approaches the size of the parent system. \response{This, however, is not the case for the SYK model in the large-$N$ limit, as discussed below.} 
\subsection{SYK model: R\'{e}nyi entropy of a non-Fermi liquid}

We now move on to large-$N$ interacting Hamiltonians, starting with the original $q=2$ SYK model described in \Eqn{eq:SYKqham}. The model connects fermions, on $N$ sites, via all-to-all two-body interactions with random complex amplitudes having variance $J^2$, which we set to unity for this section. There have been several numerical and analytical works on entanglement entropy \cite{Fu2016,Gu2017,LiuPRB2018,Huang2019,Chen2019,Zhang2020} in the SYK model for various different contexts. The analytical approaches have used standard replica path-integral appraoch for R\'{e}nyi entropy. Among these works, Gu et al. have studied growth of bi-partite R\'{e}nyi entanglement entropy in a chain of SYK dots starting from a thermofield double state \cite{Gu2017}. \response{They made a diagonal approximation in the entanglement replica space for weak inter-dot coupling. The diagonal approximation was justified based on the argument that the replica off-diagonal terms were perturbatively small in the inter-dot coupling. In our work, we find the (entanglement) replica off-diagonal terms to be very important for the R\'{e}nyi entropy for a subsystem within a single dot, since the intra-dot coupling, of course, cannot be treated perturbatively. In Sec. \ref{subsec:HigherD}, for the higher dimensional models, we consider strong inter-dot coupling and replica off-diagonal terms are also substantial there. Ref.\onlinecite{Chen2019} has used similar approximation, as in ref.\onlinecite{Gu2017},} to look into ground-state entanglement between two quadratically coupled SYK dots. During the preparation of our manuscript, we became aware of  a very recent work \cite{Zhang2020} on subsystem R\'{e}nyi entropy in the thermal ensemble for the SYK model using standard approach with the boundary conditions. Overall, their results compares well with our results on SYK model obtained using the new field-theoretic method developed in this work. 

As in the non-interacting large-$N$ model case, we compare the \response{R\'{e}nyi entropy} obtained via large-$N$ thermal field theory with that obtained via ED. However, the R\'{e}nyi entropy cannot be calculated in the interacting case using the correlation matrix approach.
\begin{figure}
    \centering
    \includegraphics[width=0.49\textwidth]{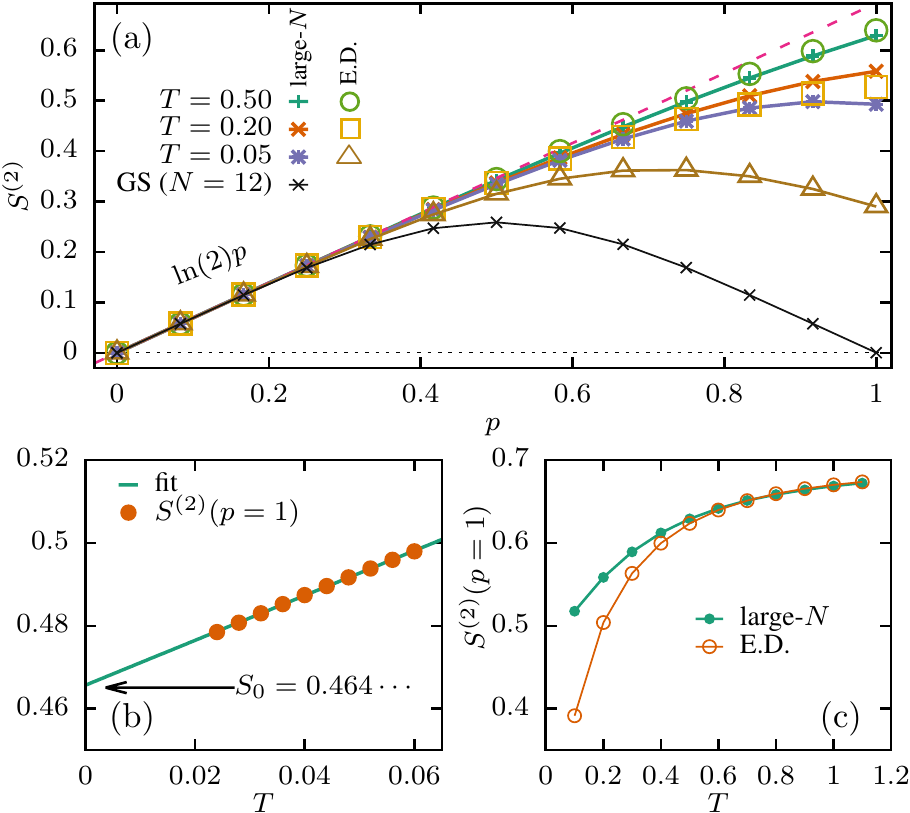}
    \caption{{\bf R\'{e}nyi entropy in SYK model:} (a) The second-R\'{e}nyi entropy ($\retwo$) for the SYK ($q=2$) model as a function of subsystem size $p$ for temperature values $T=0.50,0.20,0.05$. Predictions from large-$N$ formalism (points with lines) match exactly with ED calculations (represented by circles, squares and triangles) for subsystem sizes $p<0.4$ (both showing a volume-law scaling with a $\ln(2)$ coefficient), but deviate significantly at larger $p$ values due to the influence of residual-thermal entropy $S_0$ of the SYK model. The $\retwo$ vs. $p$ curve (black line with crosses) for the ground-state of the model, obtained using ED, is also shown for comparison. (b) R\'{e}nyi entropy for $p=1$, i.e. $\retwo(p=1)$, approaches $S_0$ linearly when temperature $T$ goes to $\emph{zero}$, as seen explicitly from a linear fit (line) to the large-$N$ data (circles) and recovering an intercept equal to $S_0\approx 0.464$. (c) The match between large-$N$ prediction (filled circles) and ED (empty circles), for $\retwo(p=1)$, is recovered at higher temperatures, see text for explanation. All ED calculations were done for $N=12$ sites.}
    \label{fig:SYK-large-N}
\end{figure}
\Fig{fig:SYK-large-N}(a) gives the $2$-nd R\'{e}nyi entropy with subsystem size $p$ for several values of $T$. We find that, yet again, for all temperature values the results from large-$N$ and ED calculations agree really well, when $p<0.4$, even for exact diagonalization performed for $N=12$ sites. Also as expected, volume law with a $\ln(2)$ coefficient still holds as $p\to 0$ for all temperature values. Interestingly though, as $T\to 0$, the large-$N$ result deviates substantially from the ED result when $p\geq 0.4$ and especially when $p\to 1$. While the \response{R\'{e}nyi entanglement entropy for the ground state}, obtained by ED, goes to \emph{zero} as expected, the large-$N$ result seems to converge linearly to a finite value when $T$ is reduced as shown in \Fig{fig:SYK-large-N}(b), where we plot $S^{(2)}(p=1)$ for low values of $T$. In fact, when extrapolated to $T\to 0$, we find the intercept to be equal (within numerical accuracy) to the thermal residual entropy $S_0\approx 0.464\cdots$ of the SYK model. \response{Remarkably, the residual entropy has found a way to influence the zero-temperature R\'{e}nyi entropy of the SYK model in the large-$N$ limit.} An explanation is provided when we look back at the expression for R\'{e}nyi entropy in \Eqn{eq:S2pbetadef}. Since by definition $\lim_{p\to1}F_{EE}(p,\beta)=F(2\beta)$, expanding the entanglement free energy to first order in $T$ ($=\beta^{-1}$) we find
\begin{align}\mylabel{eq:S0-to-EE}
    S^{(2)}(p\to1)=&\lim_{T\to 0}\frac{F(T/2)-F(T)}{T/2}=\lim_{T\to 0}\frac{\frac{T}{2}F(0)-T F(0)}{T/2}\notag\\
    =&-\left.\frac{\del F}{\del T}\right|_{T=0}=S_{0},
\end{align}
i.e. $S^{(2)}(p=1)$ is indeed equal to the residual entropy $S_0$. The above analysis suggests that the limit $N\to\infty$ and $T\to 0$ do not commute. It is established that the residual entropy for the SYK model is a consequence of exponentially small in $N$ level spacings arising from the large-$N$ limit. Taking the large-$N$ limit first also prohibits ``direct" access to the ground-state quantum entanglement when temperature is reduced, since any $T$ however small cannot resolve these exponentially small level spacings. We therefore argue that temperatures higher than the typical level spacings obtained from ED should ``mimic" this large-$N$ effect and predictions from large-$N$ thermal field theory should match with finite-$N$ ED calculations. To test this hypothesis, we plot the $p=1$ value for $2$-nd R\'{e}nyi entropy obtained from large-$N$ and finite-$N$ ED calculations for higher values of temperature in \Fig{fig:SYK-large-N}(c). Indeed, we find that the results match perfectly for $T\geq 0.5$, thereby validating our intuition about the role of residual entropy in \response{$T=0$ R\'{e}nyi entropy}  at large-$N$.
\subsection{SYK model with disordered hopping: R\'{e}nyi entropy of an interacting Fermi liquid}
We now discuss the behavior of entanglement in an interacting Fermi liquid, like the one described by the Hamiltonian in \Eqn{eq:SYKFLham}. The addition of a $q=1$ quadratic term $t_{ij}\cd_i c_j$ has been shown \mycite{Fu2016,BanerjeeAltman2016} to change the SYK NFL ground state to an interacting zero-dimensional Fermi-liquid.
\begin{figure}
    \centering
    \includegraphics[width=0.49\textwidth]{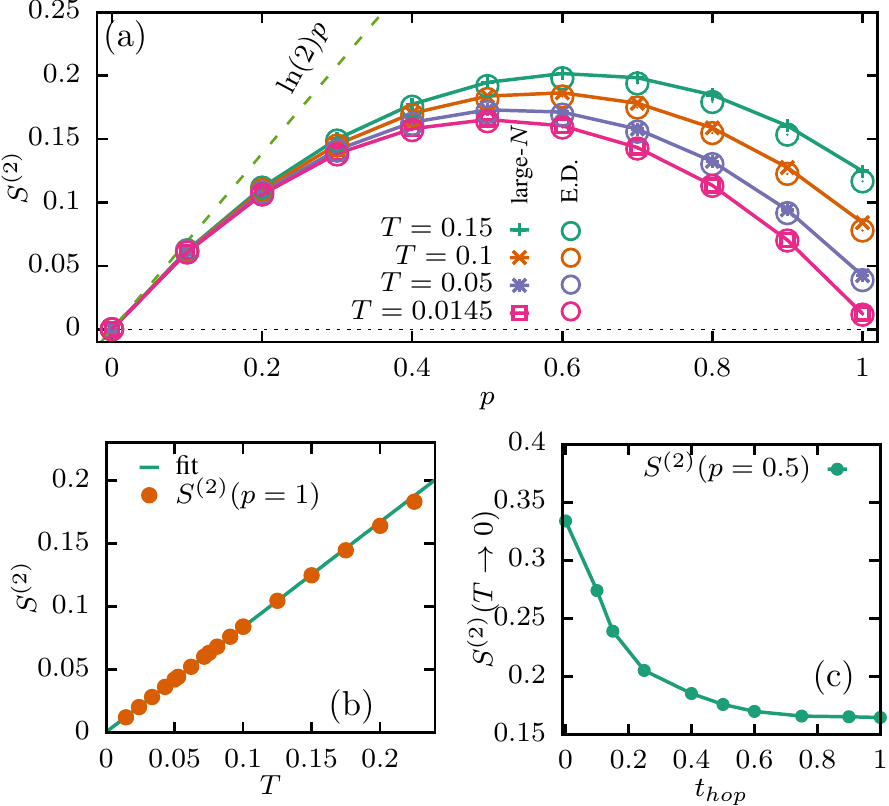}
    \caption{{\bf R\'{e}nyi entropy of interacting Fermi-liquid:} (a) Subsystem size ($p$) dependence of second R\'{e}nyi entropy, $\retwo$, in the zero-dimensional Fermi-liquid model (defined in \Eqn{eq:SYKFLham}) with SYK interactions, for several values of temperature $T$. The strength of hopping -- $\thop$ and interactions -- $J$ are both set to \emph{one}. The predictions from the large-$N$ formalism (points) and results from exact diagonalisation (circles) match for all values of $p$ and $T$. Both predict a volume-law scaling with a $\ln(2)$ coefficient. (b) The R\'{e}nyi entropy for $p=1$ (represented by circles) shows a linear dependence on $T$ (for values less than $T=0.15$) and goes to \emph{zero} as $T\to 0$, indicating an approach to a ground-state with \emph{no} residual-thermodynamic entropy. (c) The zero-temperature bipartite second-R\'{e}nyi entropy, $\retwo(p=0.5,T\to 0)$, shown as a function of increasing hopping strength $\thop$ ($J=1$), with $\thop=0$ being the SYK-NFL limit. As hopping increases, \response{bipartite R\'{e}nyi entropy} decrease from a rather large value ($\approx 0.35$) and saturates to a value ($\approx 0.15$) close to that of the non-interacting model (see \Fig{fig:nI-large-N}).}
    \label{fig:FL-large-N}
\end{figure}
 We set $\thop=1$, $J=1$ for the variances of $t_{ij}$, $J_{ij}$ in \Eqn{eq:SYKFLham} and compare the $2$-nd R\'{e}nyi entropy obtained from large-$N$ theory with ED in \Fig{fig:FL-large-N}(a). Quite encouragingly, we find an excellent match between the two for all values of temperatures! As expected, the volume law for entanglement is shown to hold as $p\to 0$ due to the all-to-all nature of interactions and hoppings. Interestingly this time, unlike the SYK NFL case, the entanglement entropy for $p=1$ goes to \emph{zero} as temperature approaches \emph{zero} (see \Fig{fig:FL-large-N}(b)). Looking back at the analysis in \Eqn{eq:S0-to-EE}, $S^{(2)}(p\to1)\to0$ implies that the FL ground state does not possess any residual entropy. Indeed, we find in literature \mycite{FuThesis2018} that the presence of a quadratic term in \Eqn{eq:SYKFLham} drastically reduces the exponentially large density of levels near the ground state thereby reducing the residual entropy to \emph{zero}. 
 
 It is interesting to explore how entanglement entropy changes as the relative strength of interactions is increased with respect to the quadratic term, going from a non-interacting system to a NFL, via heavy Fermi liquid states for $\thop\ll J$. To this end, we set $J=1$ and track the \emph{bipartite R\'{e}nyi entropy}, i.e. $S^{(2)}(p=1/2)$, as the value of $\thop$ is increased from $0$ to $1$. The bipartite R\'{e}nyi entropy is the maximal value that can be attained for any homogeneous system. \Fig{fig:FL-large-N}(c) shows the result of this exercise. When $\thop=0$, the model is described by purely SYK type interactions and therefore bipartite R\'{e}nyi entropy attains the maximum possible value $\approx 0.5\ln(2)$. Interestingly, as $t$ is increased the entanglement decreases continuously, indicating that a FL of this type becomes less entangled when the relative strength of interactions is reduced. Finally, for values of $\thop\geq J$ the bipartite entanglement entropy saturates to around $0.16$, a value close to that of non-interacting large-$N$ model (see \Fig{fig:nI-large-N}), implying that interactions become irrelevant in this limit.

\subsection{BA model: R\'{e}nyi entropy across a non-Fermi liquid to Fermi liquid transition}
One of the important aspects of the putative duality between SYK model and black holes in quantum gravity \cite{KitaevKITP,Kitaev2018,Sachdev2015,Maldacena2016} is the residual entropy and its possible connection with the black hole entropy. Hence, it is interesting to ascertain if there is any relation between the residual entropy and the ground-state entanglement entropy of SYK NFL. We saw that the residual entropy inevitably appears in the R\'{e}nyi entropy when the large-$N$ limit is taken followed by the $T\to 0$ limit in the pure SYK model. Here we discuss the BA model \cite{BanerjeeAltman2016} which helps us to make the connection between the residual and \response{$T=0$ R\'{e}nyi entropy} more explicit by tuning a QPT between a SYK NFL and a FL. 

The BA model is defined in \Eqn{eq:BAModel}. A $T=0$ transition between the chaotic NFL fixed point and non-chaotic FL fixed point is achieved by tuning the ratio of sites $p=N_\psi/N_c$. 
The quantum-critical point (QPT) occurs at $p=1$, below which the $c$-fermions behave as a SYK-NFL and the whole system has a finite $p$-dependent residual entropy, $S_0(p)=[(1-p)/(1+p)]S_0$, that goes to \emph{zero} at the critical point \cite{BanerjeeAltman2016}. Motivated by this, we ask whether any such sharp features exists in the R\'{e}nyi entropy. We analyze the $2$-nd R\'{e}nyi entropy for the model at the large-$N$ saddle point (see \appcite{app:OtherModels}) using the following two choices of subsystem $A$ -- (a) when $A$ is made up of all the $c$-fermions that have the SYK type interactions, and (b) when the $A$ is comprised of the all the non-interaction $\psi$-fermions. For our calculations we set $J=\thop=V=1$ (see \Eqn{eq:BAModel}). 

\begin{figure}
    \centering
    \includegraphics[width=0.49\textwidth]{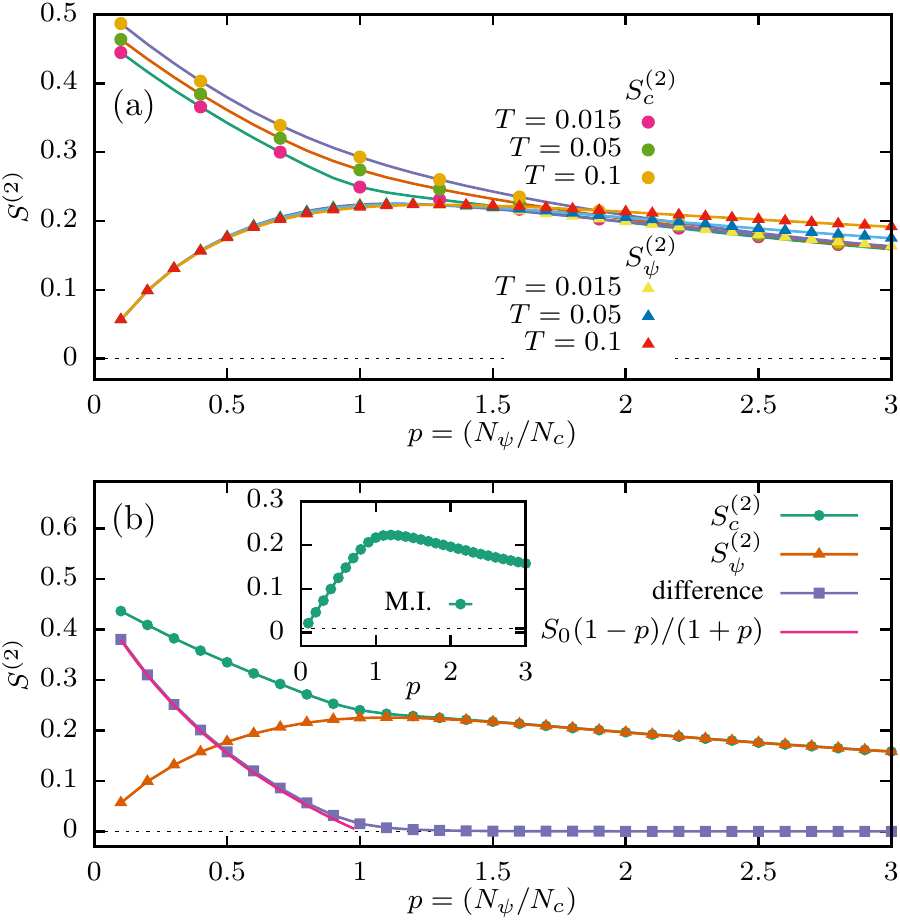}
    \caption{{\bf R\'{e}nyi entropy in BA model:} (a) Subsystem $2$-nd R\'{e}nyi-entropy for the $c$-fermions -- $\retwo_c$ (circles with lines) and $\psi$-fermions -- $\retwo_\psi$ (triangles), for the BA model (\Eqn{eq:BAModel}), shown as function of site fraction $p=N_\psi/N_c$ for temperatures $T=0.015,0.05,0.1$. (b) Extrapolated zero-temperature R\'{e}nyi-entropies, $\retwo_c$ (circles) and $\retwo_\psi$ (triangles), showing the signature of the underlying QPT. $\retwo_c$ and $\retwo_\psi$ become exactly equal at the critical point for the QPT, $p=1$, an remain equal for $p>1$. Inset, gives the mutual information (M.I.) between the $c$ and $\psi$ fermions showing a peak at the critical point. Furthermore, the difference of the R\'{e}nyi-entropies, $\retwo_c-\retwo_\psi$ (squares with lines), precisely matches the analytical prediction, $S_0(p)=S_0(1-p)/(1+p)$ (solid line), for the $p$-dependent residual entropy for the BA model from ref.\onlinecite{BanerjeeAltman2016}.}
    \label{fig:BA-model}
\end{figure}

The subsystem-R\'{e}nyi entropy for the $c$-fermions, $\retwo_c$ (case (a)), and $\psi$-fermions, $\retwo_\psi$ (case (b)), are shown as a function of site fraction $p$ in \Fig{fig:BA-model}(a) for multiple values of temperature $T$. Ideally, at $T=0$, in a pure state one expects $\retwo_c=\retwo_\psi$. However, yet again, we find that residual entropy $S_0(p)$ has managed to influence entanglement entropy as seen from $\retwo_c$ (represented by circles) approaching a value close to $S_0=0.464$ of the SYK model when $p\to0$. On the other hand, the $\retwo_\psi$ (triangles) goes to \emph{zero} in the same limit in accordance with the volume-law maximal entanglement, $p\ln(2)$. To uncover the signature of the underlying QPT, we extrapolated $\retwo_c$ and $\retwo_\psi$ to $T\to0$ and plot the results in \Fig{fig:BA-model}(b). Remarkably, we find that around the critical point, $p=1$, the R\'{e}nyi-entropy for the fermions become exactly equal , i.e., $\retwo_c=\retwo_\psi$ and remains equal for all values $p>1$.  Motivated by this, to further understand the relationship between the residual entropy $S_0(p)$ and the R\'{e}nyi-entropies $\retwo_c$, $\retwo_\psi$, we compare $S_0(p)$ with the difference $\retwo_c-\retwo_\psi$, which approaches $S_0$ for $p\to 0$ and also vanishes continuously at $p=1$. Quite encouragingly, we find the match to be excellent, as seen in \Fig{fig:BA-model}(b). Thus the difference between the R\'{e}nyi entropies of the larger and smaller subsystems directly corresponds to the residual entropy and carries the signature of the underlying QPT. 
\response{The preceding results also leads us to conjecture that the zero-temperature R\'{e}nyi entropy of the BA model has contributions from both the exponentially dense energy levels and the true ground-state entanglement for $p\leq 1$. In particular, the results suggest that the true ground-state entanglement primarily determines the R\'{e}nyi entropy for the $\psi$-fermions -- $\retwo_\psi$, while the large many-body density of states, arising from the large-$N$ limit, affects the R\'{e}nyi entropy for $c$-fermions -- $\retwo_c$. Further exploration along this direction should prove useful in separating the two contributions}. 
We also look into the mutual information (per site) between the two species of fermions, defined as $I(p)={1\over(1+p)}\retwo_c+{p\over(1+p)}\retwo_c-\retwo_{c\psi}$. $I(p)$ has a broad peak around the critical point as shown in \Fig{fig:BA-model}(b) inset.  Indeed, the underlying QPT has left an imprint in the entanglement entropy.

\subsection{Higher dimensional large-$N$ models: R\'{e}nyi entropy in an interacting diffusive metal} \label{subsec:HigherD}
We now discuss one of the main new results obtained using the path-integral formalism developed in this work. We compute the R\'{e}nyi entropy of a sub-region in an interacting diffusive metal. To this end, we consider the higher-dimensional generalizations of the SYK derived large-$N$ models, which were discussed in \Seccite{sec:allmodels} and can be described by a Hamiltonian with the general form given in \Eqn{eq:HD-SYK-chain}.  The large-$N$ equations needed to determine \response{R\'{e}nyi} entropy were derived in \Eqn{eq:large-N-1D-SP} for one-dimensional large-$N$ models. Here we quantify how well our the formalism performs in predicting the R\'{e}nyi entropy for these systems. 
\begin{figure*}
    \centering
    \includegraphics[width=\textwidth]{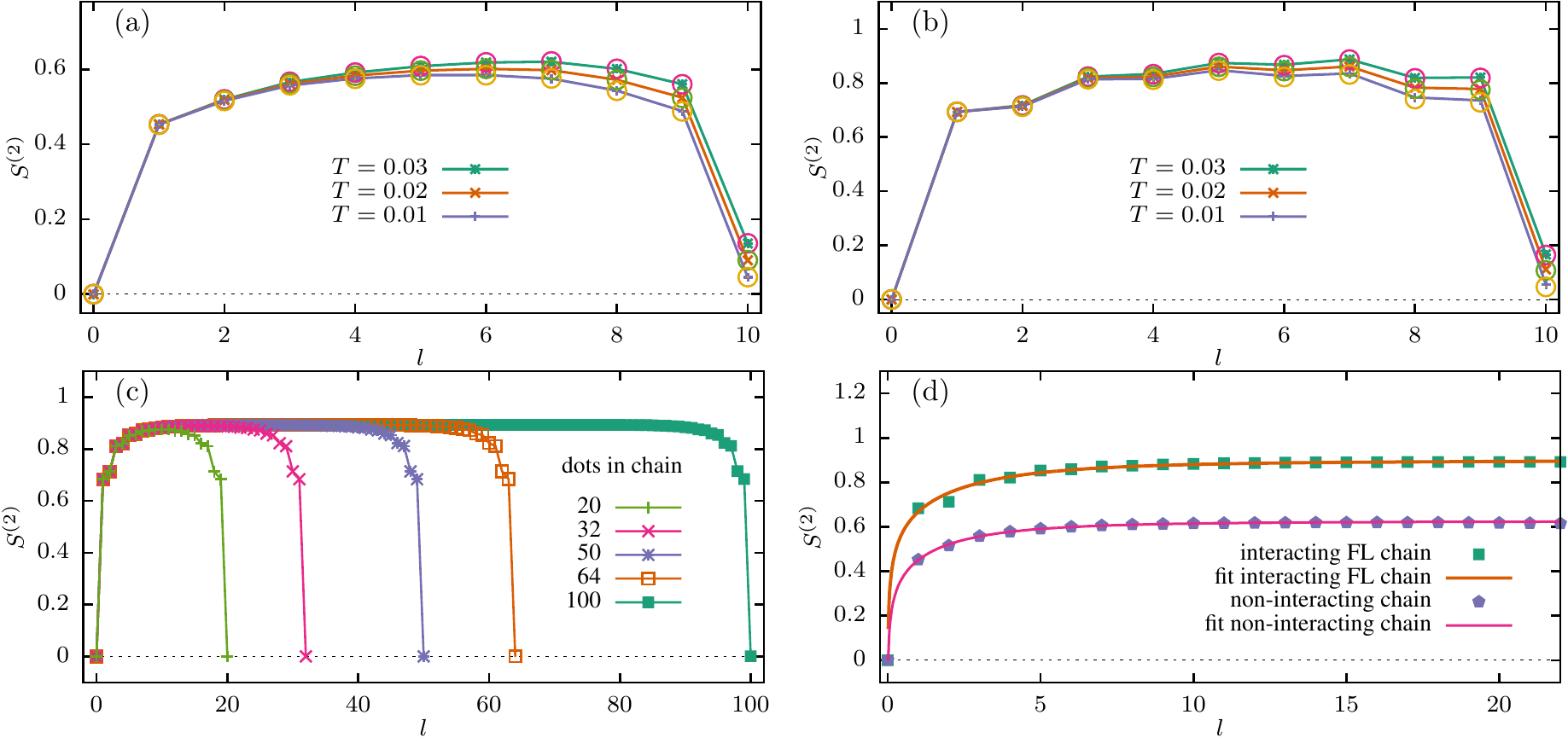}
    \caption{{\bf R\'{e}nyi entropy of diffusive Fermi liquid:} (a) Behavior of R\'{e}nyi entropy, $\retwo$, vs. subsystem size $l$  in a model of non-interacting dots coupled to nearest neighbors in a ring of $N_{dots}=10$ dots (see \Fig{fig:SYK-models}(d)), for temperatures $T=0.03,0.02,0.01$. The model is obtained by setting $q=q_{hop}=1$ and  intra as well as inter-dot hopping strengths $\tchain=J=1$  in \Eqn{eq:HD-SYK-chain}. (b) R\'{e}nyi entropy vs. $l$ for the non-interacting model when $\tchain=1$, $J=0$. In both (a) and (b), large-$N$ predictions (points joined by lines)  are in agreement with the results obtained from exact numerics (color matched circles) performed using correlation matrices. (c) Extrapolated zero-temperature R\'{e}nyi entropy as a function of subsystem size $l$ for a ring of SYK dots, (\Fig{fig:SYK-models}(d)), connected by random hoppings. We set the interaction strength $J=1$ within each SYK dot as well as  the strength of hopping between dots $\tchain=1$.  The number of dots in the chain, $N_{dots}$, is varied from $20-100$ dots (well beyond the scope of any exact numerics), to verify that the growth of entanglement for small subsystem sizes becomes independent of $N_{dots}$ for $N_{dots}>20$. (d) A close up, showing the growth of $\retwo$ near small subsystem sizes, i.e. $l\leq 20$, in the interacting SYK-chain (represented by squares) and the non-interacting chain (represented by pentagon) discussed in (a), for a large value of chain size $N_{dots}=50$. Both curves can be fitted (lines) with the analytical function $S^{(2)}\sim \log(1/(l^{-2}+\mfp^{-2})^{1/2})$, where $\mfp$ is an emergent ``mean-free path" length scale which takes a value between $\sim 4-5$ in this case.}
    \label{fig:FL-large-N-extended}
\end{figure*}
As mentioned earlier (see the discussion below \Eqn{eq:large-N-1D-SP}), naturally solving the saddle-point equations for the extended system of the one-dimensional chain is computationally more expensive than the zero-dimensional case since we need to retain the Green's function $G^{(x)}_{\sigma\sigma'}(\tau_1,\tau_2)$ for each  SYK dot in the chain. In particular, the time complexity for solving the equations scales linearly with the number of dots which we denote as $N_{dots}$. However, this increase in time complexity does not prevent us from accessing entanglement for large chains and we are able to perform calculations for values of $N_{dots}$ which are well beyond the reach of ED, even for the non-interacting large-$N$ models for which we use the correlation matrix approach of \Seccite{subsec:nI-formalism}. Therefore, we adopt the following strategy -- first, we compare the results of our large-$N$ field-theoretic approach with that obtained from correlation matrix for non-interacting systems for moderate system sizes, and then explore the behavior of entanglement for large non-interacting and interacting systems. The interacting systems are much larger than that accessible via ED.

We set $q_{hop}=q=1$ in \Eqn{eq:HD-SYK-chain}, making the model non-interacting, and look at two configurations of hopping amplitudes $\tchain=J=1$ and $\tchain=1, J=0$. Also, we set $N_{dots}=10$ and $N=10$, i.e. a moderate number of flavors per dot to perform exact numerics using the correlation matrix approach of \Seccite{subsec:nI-formalism}. This allows us to access subsystems sizes for $p=0.0-1.0$ in steps of $0.1$. The result for the above two scenarios are shown in \Fig{fig:FL-large-N-extended}(a) and (b) respectively. We find, like before, an \emph{excellent} agreement between large-$N$ predictions and exact numerical calculations, with the results for R\'{e}nyi entropy from both the techniques collapsing on top of each other for multiple temperature values. 

Having established the agreement of our formalism with exact numerics, we move on to larger chain sizes involving interacting as well as non-interacting dots. In particular, we keep $q_{hop}=1$ and study chains with either $q=2$ or $q=1$ body interaction for the dots (see \Eqn{eq:HD-SYK-chain}). The resultant ground state in these cases is known to describe either a strongly-interacting diffusive Fermi-liquid \mycite{Song2017} or a non-interacting (but still diffusive) FL, both of which are gapless phases. The entanglement scaling with subsystem size ($l$) for a FL have been argued to violate the area law. For e.g., in 1-d translationally invariant gapless Fermi system, conformal field theory (CFT) predicts \mycite{Calabrese2009} that the von-Neuman entanglement entropy $S^{(1)}$ (obtained by taking the limit $\lim_{n\to 1}S^{(n)}$), scales with the universal form $S^{(1)}=(c/3)\ln(l)+const.$  where $c$ is the central-charge of the underlying CFT. Following which, an exact expression for $S^{(1)}$, of the form $S^{(1)}\sim l^{d-1}\log{l}$,  was obtained for non-interacting FLs in arbitrary dimensions $d$ \mycite{GioevPRL2006}. Later on, it was argued that the $n$-th order R\'{e}nyi entropy for non-interacting as well as interacting FLs should follow $S^{(n)}={1\over2}(1+{1\over n})S^{(1)}$, and therefore should have the same scaling with system size \mycite{SwinglePRL2010,SwinglePRB2012}. The above scaling form has been studied and confirmed, for systems with and without interactions, using several numerical approaches \mycite{McminisPRB2013}. Therefore, we may naively expect to recover this $\log l$ scaling for our model of a one-dimensional chain of connected dots. 

To verify this, we evaluate the $2$-nd R\'{e}nyi entropy, in the $T\to0$ limit, as a function of subsystem size $l$ for increasing values of $N_{dots}$ and plot the result in \Fig{fig:FL-large-N-extended}(c). We find that the curves converge as the chain size increases and entanglement entropy growth near $l=0$ becomes independent of the total number of dots in the chain.
Surprisingly, we find that R\'{e}nyi entropy quickly saturates to a finite value as subsystem size is increased, a behavior \emph{inconsistent} with $\log l$ scaling, which poorly fits the the entanglement growth curve in our calculation. Instead, we find that a ``modified" growth function
\begin{align}\mylabel{eq:EE_growth_xi}
    S^{(2)}(l)\sim \log\left[\frac{1}{\sqrt{l^{-2}+\mfp^{-2}}}\right]+\mathrm{const.},
\end{align}
fits our data rather well, see \Fig{fig:FL-large-N-extended}(d). This indicates the presence of an emergent length scale $\mfp$  in the gapless system! This could be explained by realizing that the FL state is obtained by connecting sites (dots) having random intra-dot couplings, which are uncorrelated at different sites, with random hoping amplitudes (for e.g. $t_{ijxx'}$ in \Eqn{eq:HD-SYK-chain}). The former will induce an effective mean free path for the quasi-particles in the system and therefore will cut off the growth of entanglement beyond a length scale $l_0$. Indeed, if we identify $\mfp$ as the mean-free path and take $\mfp\to\infty$, we recover the expected $\log l$ growth of entanglement. However, we note that a definition of mean free path in a system with random inter-site hopping (with zero mean) is somewhat subtle due to the difficulty in identifying a `Fermi velocity'. Interestingly, the growth function in \Eqn{eq:EE_growth_xi} has also been suggested in ref.\onlinecite{potter2014} based on numerical studies of system size scaling of entanglement in a disordered non-interacting system, and hydrodynamic arguments for diffusive FL. On the contrary, our results explicitly demonstrate the scaling law of \Eqn{eq:EE_growth_xi} in an interacting diffusive metal for system sizes much beyond any other numerical techniques. The coefficient of the growth function in \Eqn{eq:EE_growth_xi} is also expected to contain the information of effective central charge \cite{potter2014} of the underlying CFT. As evident in \Fig{fig:FL-large-N-extended}(d), the coefficient of the growth function changes with interaction strength. These aspects will be studied in detail in a future work \cite{Haldar2020Renyi}.

\section{Conclusions and discussions}
In this work, we have proposed and developed a new equilibrium and non-equilibrium field theory method to compute R\'{e}nyi entanglement entropies for interacting fermions. The basis of the field-theory formalism relies on an operator identity that we have derived here. The path-integral technique is an alternative, and maybe complementary, to the existing path-integral methods that typically require complicated boundary conditions on the fields for computing R\'{e}nyi entropy. Our method rigorously transforms the complex boundary conditions into time-dependent self-energies while preserving the familiar anti-periodic boundary conditions on the fermionic fields as in a usual coherent-state path integral. As a result, the path-integral formalism could be easily incorporated within standard weak-coupling diagrammatic field-theory techniques and approximations, e.g. mean-field theory and random-phase approximation (RPA), in a very transparent manner. The path integral technique to compute entanglement entropy could also be integrated with strong-coupling approaches, e.g. in Hubbard model, to go beyond Gaussian actions \cite{Ghosh2019}. We demonstrate this by formulating the field theory for R\'{e}nyi entropy in Hubbard model within DMFT approximation, and in several interacting large-$N$ fermion models of current interest. 

Using the formalism, we obtain several important results on \response{R\'{e}nyi} entropy of FL and NFL states in zero-dimensional interacting large-$N$ models, i.e. SYK and related models. We exactly compute the second R\'{e}nyi entropy of a subsystem as a function relative subsystem size $p$ in the $N\to\infty$ limit. We analytically show that the subsystem is maximally entangled with an entanglement entropy $p\ln(2)$ at half filling in the limit $p\to 0$. By comparing with exact diagonalization result for small system sizes $N=10-12$, we demonstrate that the $1/N$ corrections are very small for entanglement entropy for non-interacting systems as well as for strongly interacting heavy Fermi liquids. However, the $1/N$ corrections are found to be large and non-perturbative for $p\gtrsim0.4$ in the SYK model, where the residual entropy of the NFL state contributes to the $T\to 0$ R\'{e}nyi entropy in the large-$N$ limit, as we analytically show for $p\to 1$. Thus, our results reveal intriguing connections between residual entropy of SYK NFL and its $T\to 0$ R\'{e}nyi entropy in the large-$N$ limit, calling for a proper interpretation of \emph{residual} R\'{e}nyi entropy, in addition to the pure ground-state quantum entanglement, for a general $p$. We further make the connection between residual entropy and R\'{e}nyi entropy explicit in the BA model where the difference of $T\to 0$ R\'{e}nyi entropy of larger and smaller subsystems, for a particular subdivision of the system, can be directly attributed to the residual entropy of the NFL state. Finally, using our method, we obtain non-trivial system-size scaling of entanglement in an interacting diffusive metal for system sizes much beyond that accessible via exact diagonalization methods. We find the existence of an emergent length scale that limits the growth of entanglement entropy in the diffusive metal. 

The DMFT formulation developed here for R\'{e}nyi entropy would be very useful to understand quantum entanglement in correlated systems. The formulation could be easily integrated with standard impurity solver, e.g. CTQMC \cite{Gull2011}. Entanglement entropy has been previously studied within CTQMC treating the interaction correction perturbatively for weak interaction \cite{Wang2014}. Our method is non-perturbative and can be used in the strongly interacting regime. The trace formula and the path integral derived here might also be useful to compute R\'{e}ny entropy of fermions in other QMC techniques, like determinantal QMC (DQMC) \cite{Assaad2008,Hastings2010,Humeniuk2012,Grover2013,Wang2014}.

The field theory developed in this work is also amenable to perturbative renormalization group (RG) methods, e.g. applied to non-equilibrium situation \cite{Sarkar2014}, that may be  advantageous for deriving analytical results on entanglement entropy in interacting fermionic systems. It will be interesting to explore possible recursion relation between R\'{e}nyi entropies at different orders ($n$) to understand the behaviour as a function of $n$ and take appropriate limit to compute von Neumann entanglement entropy and entanglement negativity \cite{Vidal2002,Calabrese2012,Lee2013,Wu2019,Lu2019} for interacting fermions. For the large-$N$ fermionic models, it will be desirable to go beyond the $N\to \infty$ limit to compute fluctuations around the saddle-point of the entanglement action, e.g. by generalizing the methods of ref.\onlinecite{Bagrets2016} for SYK thermal field theory. The time-dependent self-energy kick in the real-time action (e.g. in \Eqn{eqn:EE-K-final-def}) can be mimicked by time-dependent non-Hermitian terms in a Hamiltonian or non-unitary terms in the time evolution. Hence, a protocol, inspired by the ``kick" interpretation, may also be designed to measure the R\'{e}nyi-entropy in quantum circuits.
\begin{acknowledgements}
We thank Rajdeep Sensarma and Saranyo Moitro for useful discussions and for collaboration at the initial stages of the work. We also thank Thomas Scaffidi for discussions between him and AH at the final stages of this work. SB acknowledges support from The Infosys Foundation (India), and SERB (DST, India) ECR award. AH acknowledges Compute Canada for the use of their cluster framework for performing the numerical computations presented in this paper.
\end{acknowledgements}

\appendix
\section{Operator expansion} \label{app:OperatorExpansion}
In this appendix we derive the operator expansion identity of \Eqn{eq:op-exp-Dn-Dn}. The displacement operator defined in \Seccite{sec:formulation} can be used to represent \mycite{Cahill1999} the Dirac delta function for Grassmann numbers
\begin{align}\mylabel{eq:delta-grassman-def-2}
    \delta(\bs{\xi}-\bs{\eta})=\Tr[D_{N}(\bs{\xi})E_{A}(\bs{\eta})],
\end{align}
where the operator $E_A$ is defined as
\begin{align}\mylabel{eq:EA-def}
    E_{A}(\bs{\eta})=\int d^{2}\bs{\alpha}\ \exp\left[\sum_{i}\eta_{i}\bar{\alpha}_{i}-\alpha_{i}\bar{\eta}_{i}\right]
    |\bs{\alpha}\rangle\langle-\bs{\alpha}|.
\end{align}
The representation of the delta function (\Eqn{eq:delta-grassman-def-2}) allows the displacement operators to
form a basis for expanding an arbitrary operator $F$, such that
\begin{align}\mylabel{eq:op-exp-1}
    F=\int d^{2}\bs{\xi}\ \Tr[FE_{A}(\bs{\xi})]D_{N}(\bs{\xi}),
\end{align}
Although the operator expansion formula in \Eqn{eq:op-exp-1} is useful, it has certain drawbacks. In particular, \Eqn{eq:op-exp-1} is not amenable to a path-integral formulation with \emph{simple} boundary conditions mainly because of the outer product $|\bs{\alpha}\rangle\langle-\bs{\alpha}|$ in the definition of $E_A$ (\Eqn{eq:EA-def}). Therefore, a generalization of \Eqn{eq:op-exp-1} involving only the displacement operators $D_N$ is desirable. In this context, an insightful observation is to realize that even the operator $E_A$ can be decomposed using \Eqn{eq:op-exp-1}, as shown below
\begin{align}
    E_{A}(\bs{\eta})=\int d^{2}\bs{\xi}\ \Tr[E_{A}(\bs{\eta})E_{A}(\bs{\xi})]D_{N}(\bs{\xi}),
\end{align}
which can then be used to transform \Eqn{eq:op-exp-1} into
\begin{align}\mylabel{eq:op-exp-Dn-Dn-app}
    F=\int d^{2}\bs{\xi},\bs{\eta}\ f_{N}(\bs{\eta},\bs{\xi})\Tr[FD_{N}(\bs{\xi})]D_{N}(\bs{\eta}),
\end{align}
where
\begin{align}\mylabel{eq:fn-gamma-xi-def}
    &f_{N}(\bs{\gamma},\bs{\xi})\equiv\Tr[E_{A}(\bs{\gamma})E_{A}(\bs{\xi})]\notag\\
    &=2^{N}\exp\left[\sum_{i}-\frac{1}{2}(\bar{\gamma_{i}}\gamma_{i}+\bar{\xi}_{i}\xi_{i})\right]\exp\left[\sum_{i}\frac{1}{2}(\bar{\gamma_{i}}\xi_{i}-\bar{\xi}_{i}\gamma_{i})\right].
\end{align}
\Eqn{eq:op-exp-Dn-Dn-app} offers a way to decompose a general operator $F$ using \emph{only} the normal-ordered displacement operators $D_N$. Using the relation between the usual displacement operators and their normal-ordered counterparts in \Eqn{eq:Dn-def}, an equivalent identity involving the operator $D(\bs{\xi})$ can be derived, i.e.,
\begin{align}\mylabel{eq:op-exp-D-D}
F=\int d^{2}\bs{\xi},\bs{\eta}\ f(\bs{\eta},\bs{\xi})\Tr[FD(\bs{\xi})]D(\bs{\eta}),
\end{align}
where
\begin{align}\mylabel{eq:f-xi-eta-def}
    f(\bs{\eta},\bs{\xi})	=&2^{N}\exp\left[\sum_{i}\frac{1}{2}(\bar{\eta_{i}}\xi_{i}-\bar{\xi}_{i}\eta_{i})\right]\notag\\
	=&\exp\left[\sum_{i}\frac{1}{2}(\bar{\eta_{i}}\eta_{i}+\bar{\xi}_{i}\xi_{i})\right]f_{N}(\bs{\eta},\bs{\xi}).
\end{align} 
Both the functions $f$, defined above, and $f_N$ (see \Eqn{eq:fn-gamma-xi-def}) satisfy
\begin{align}\mylabel{eq:f-fN-prop}
    f_{N}(-\bs{\eta},\bs{\xi})=f_{N}(\bs{\eta},-\bs{\xi})=f_{N}(\bs{\xi},\bs{\eta}).
\end{align}
It can be shown that cyclic rule for trace gets modified due to the presence of the displacement operators, i.e.
\begin{align}
\Tr[D_{N}(\bs{\eta})G] & =\Tr[GD_{N}(-\bs{\eta})]\label{eq:Tr-DN-G-main}\\
\Tr[D(\bs{\eta})G] & =\Tr[GD(-\bs{\eta})],\label{eq:Tr-D-G-main}.
\end{align}
The trace identity of \Eqn{eq:Tr-FG-exp-DN} can be obtained using Eqs.\eqref{eq:op-exp-Dn-Dn}, \eqref{eq:f-fN-prop}, and \Eqn{eq:Tr-DN-G-main}.

\newcommand{\bm}[1]{\bs{#1}}
\section{The characteristic function hierarchy} 
In this section we give the details of the steps that went into deriving the R\'{e}nyi-entropy hierarchy reported in \Seccite{subsec:nI-formalism} of the main text.
\subsubsection{Derivation}\mylabel{part:Deriving-hiearchy-details}
We use the operator expansion formula \eqn{eq:op-exp-D-D} to
write 
\begin{align}
\rho^{n+1}D(\bs{\alpha})=&\rho^{n}\rho\ D(\bs{\alpha})\notag\\
=&\int d^{2}\bs{\xi},\bs{\eta}\ f(\bs{\eta},\bs{\xi})Tr[\rho^{n}D(\bs{\xi})]D(\bs{\eta})\rho D(\bs{\alpha}).
\end{align}
Taking trace on both sides gives
\begin{equation}
\Tr[\rho^{n+1}D(\bs{\alpha})]=\int d^{2}\bs{\xi},\bs{\eta}\ f(\bs{\eta},\bs{\xi})\Tr[\rho^{n}D(\bs{\xi})]\Tr[D(\bs{\eta})\rho D(\bs{\alpha})]
\end{equation}
 Using the modified cyclic rule for trace \eqn{eq:Tr-D-G-main} and the product formula
\begin{equation}
D(\bs{\xi})D(\bs{\eta})=D(\bs{\xi}+\bs{\eta})\exp\left[\frac{1}{2}\sum_{i}\left(\bar{\eta}_{i}\xi_{i}-\bar{\xi_{i}}\eta_{i}\right)\right],\label{eq:D-prod-rule}
\end{equation}
we have 
\begin{align}
\Tr[D(\bs{\eta})\rho D(\bs{\alpha})]&=\Tr[\rho D(\bs{\alpha})D(-\bs{\eta})]\notag\\
&=\Tr[\rho D(\bs{\alpha}-\bs{\eta})]\exp[\frac{1}{2}(\sum_i\bar{\alpha}_i\eta_i-\bar{\eta}_i\alpha_i)]\notag\\
&=\Tr[\rho D(\bs{\alpha}-\bs{\eta})]\frac{f(-\bs{\eta},\bs{\alpha})}{2^{N_{A}}}.
\end{align}
Putting this back, we get
\begin{align}
\Tr[\rho^{n+1}D(\bs{\alpha})]=&\frac{1}{2^{N_{A}}}\int d^{2}\bs{\xi},\bs{\eta}\ f(\bs{\eta},\bs{\xi})f(-\bs{\eta},\bs{\alpha})\Tr[\rho^{n}D(\bs{\xi})]\notag\\
&\Tr[\rho D(\bs{\alpha}-\bs{\eta})],
\end{align}
 substituting $\bs{\eta}\to-\bs{\eta}$, we get \eqn{eq:recurse_1}, i.e.
\begin{align}
&\Tr[\rho^{n+1}D(\bs{\alpha})]\notag\\
 & =\frac{1}{2^{N_{A}}}\int d^{2}\bs{\xi},\bs{\eta}\ f(-\bs{\eta},\bs{\xi})f(\bs{\eta},\bs{\alpha})\Tr[\rho^{n}D(\bs{\xi})]\Tr[\rho D(\bs{\alpha}+\bs{\eta})]\nonumber \\
 & =\frac{1}{2^{N_{A}}}\int d^{2}\bs{\xi},\bs{\eta}\ f(\bs{\xi},\bs{\eta})f(\bs{\eta},\bs{\alpha})\Tr[\rho^{n}D(\bs{\xi})]\Tr[\rho D(\bs{\alpha}+\bs{\eta})],
\end{align}
 where we have used the property in \eqn{eq:f-fN-prop}.

\subsubsection{Simplification of the recursion\label{subsec:Simplification-of-the-RE-recursion}}

We continue from \eqn{eq:recurse_1}
and use the Gaussian ansatz \eqn{eq:ch-nI-gaussian-ansatz}, to
write

\begin{eqnarray*}
\chi_{n+1}(\bs{\alpha}) & = & \frac{1}{2^{N_{A}}}\int d^{2}\bs{\xi},\bs{\eta}\ f(\bs{\xi},\bs{\eta})f(\bs{\eta},\bs{\alpha})\ \chi_{n}(\bs{\xi})\chi_{1}(\bs{\alpha}+\bs{\eta})\\
 & = & \lambda_{n}\lambda\frac{1}{2^{N_{A}}}\int d^{2}\bs{\xi},\bs{\eta}\ f(\bs{\xi},\bs{\eta})f(\bs{\eta},\bs{\alpha})\ \exp[\bs{\bar{\xi}}^{T}{A_{n}}\bs{\xi}]\\
&&\exp[\bm{(\bar{\alpha}+\bar{\eta})^{T}}{A}\bm{(\alpha+\eta)}]\\
& = & 2^{N_{A}}\lambda_{n}\lambda\int d^{2}\bs{\xi},\bs{\eta}\ \exp\left[\frac{\bm{\bar{\xi}}\bs{\eta}}{2}-\frac{\bar{\eta}\bs{\xi}}{2}\right]\\
&&\exp\left[\frac{\bar{\bs{\eta}}\bs{\alpha}}{2}-\frac{\bar{\bs{\alpha}}\bs{\eta}}{2}\right]
\exp[\bm{\bar{\xi}}^{T}{A_{n}}\bm{\xi}]\\
&&\exp[\bm{(\bar{\alpha}+\bar{\eta})^{T}}A\bm{(\alpha+\eta)}]\\
 & = & 2^{N_{A}}\lambda_{n}\lambda\int d^{2}\bs{\xi},\bs{\eta}\ \exp\left[\frac{\bar{\bs{\xi}}\bs{\eta}}{2}-\frac{\bar{\bs{\eta}}\bs{\xi}}{2}\right]\\
 &&\exp\left[\frac{\bar{\bs{\eta}}\bs{\alpha}}{2}-\frac{\bar{\bs{\alpha}}\bs{\eta}}{2}\right]\\
 &&\exp[\bm{\bar{\xi}}^{T}{A_{n}}\bm{\xi}]\exp[\bm{\bar{\alpha}^{T}}A\bs{\alpha}]\\
&&\exp[\bm{\bar{\eta}^{T}}A\bs{\eta}+\bm{\bar{\alpha}^{T}}A\bs{\eta}+\bm{\bar{\eta}^{T}}A\bs{\alpha}].
\end{eqnarray*}
%
Integrating out $\bs{\xi}$ and $\bs{\eta}$, we get
\begin{align*}
 &\chi_{n+1}(\bs{\alpha})\notag\\
 = & 2^{N_{A}}\lambda_{n}\lambda\det(-A_{n})\det(-A-A_{n}^{-1}/4)\\
 &\exp\left[\bm{\bar{\alpha}^{T}}\left(A-(A-\I/2)(A+A_{n}^{-1}/4)(A+\I/2)\right)\bm{\alpha}\right].
\end{align*}
Equating the answer to the Gaussian ansatz for $n+1$, we get
\begin{align}\label{eq:route_1_a}
&\lambda_{n+1}\exp[\bm{\bar{\alpha}^{T}}{A_{n+1}}\bs{\alpha}]\notag\\
 & =2^{N_{A}}\lambda_{n}\lambda\det(-{A_{n}})\det(-{A}-{A_{n}^{-1}}/4)\notag\\
 & \exp\left[\bm{\bar{\alpha}^{T}}\left({A}-({A}-\I/2)({A}+{A_{n}^{-1}}/4)({A}+\I/2)\right)\bm{\alpha}\right].
\end{align}
Implying the following recursion relations for $\lambda_{n}$ and
$\bm{A_{n}}$
\begin{align}\mylabel{eq:recursion_An_A2}
\lambda_{n+1} & =2^{N_{A}}\lambda_{n}\lambda\det(-{A_{n}})\det(-{A}-{A_{n}^{-\I}}/4)\notag\\
&=2^{N_{A}}\lambda_{n}\lambda\det({A_{n}A}+\I/4)\notag\\
&=\det[2{A_{n}A}+\I/2]\lambda_{n}\notag\\
{A_{n+1}} & = \left({A}-({A}-\I/2)({A}+{A_{n}^{-1}}/4)({A}+\I/2)\right)\notag\\
&=(A+A_{n})/(1+4AA_{n}).
\end{align}
In the last line, we have chosen to simplify the expressions, by assuming that ${A_{n}}$ at the end will be well behaved functions of the matrix ${A}$,
and therefore commutes with ${A}$, allowing us to treat all matrices
as scalars. We rewrite the recursion for ${A_{n}}$ as
\begin{align}
A_{n+1}A & =(A^{2}+A_{n}A)/(1+4AA_{n})\nonumber \\
\implies2A_{n+1}A+\frac{1}{2} & =(2A^{2}+2A_{n}A)/(1+4AA_{n})+\frac{1}{2}\nonumber \\
\implies2A_{n+1}A+\frac{1}{2} & =\frac{(2A^{2}+2A_{n}A)+\frac{1}{2}(1+4AA_{n})}{(1+4AA_{n})}\nonumber \\
\implies2A_{n+1}A+\frac{1}{2} & =1+\frac{(A^{2}-\frac{1}{4})}{2A_{n}A+\frac{1}{2}}.\label{eq:rewrite-nI-An-rec-proof}
\end{align}
The last line gives the expression reported in the main text as \eqn{eq:final_X_recursion}.
\subsubsection{Solving the recursion}\mylabel{sec:sol-verification}
We find the following solution 
\begin{equation}
X_{n}=\left[(\I-C^T)^{n}+{(C^T)}^{n}\right]
\left[(\I-{C^T})^{n-1}+{(C^T)}^{n-1}\right]^{-1}\label{eq:Xn_rec_sol-app}
\end{equation}
solves \eqn{eq:final_X_recursion}. To prove it, we first define $B=C^T$ (and treat everything like scalars as explained earlier) to simplify the notation and then plug the above solution into the R.H.S of the recursion (\eqn{eq:final_X_recursion}) and simplify as shown below
\begin{eqnarray*}
 \bm{1}+\frac{B(B-1)}{X_{n}}
 & = & 1+\frac{B(B-1)}{\frac{(1-B)^{n}+B^{n}}{(1-B)^{n-1}+B^{n-1}}}\\
 & = & 1+\frac{B(B-1)[(1-B)^{n-1}+B^{n-1}]}{(1-B)^{n}+B^{n}}\\
 & = & \frac{(1-B)^{n}+B^{n}-B(1-B)^{n}-(1-B)B^{n}}{(1-B)^{n}+B^{n}}\\
 & = & \frac{(1-B)^{n+1}+B^{n+1}}{(1-B)^{n}+B^{n}}=X_{n+1},
\end{eqnarray*}
to find the answer to be $X_{n+1}$. Therefore, \eqn{eq:Xn_rec_sol-app}
solves the recursion in \eqn{eq:final_X_recursion}.

\section{Keldysh formulation for non-interacting systems}\mylabel{sec:Keldysh-nI}
For a non-interacting system undergoing non-equilibrium time evolution starting from an initial Gaussian state, e.g. described by the thermal density matrix of \eqn{eq:ThermalNonInt}, the characteristic function (eqns.\eqref{eq:chi-neq-def-2}, \eqref{eqn:SEEK-def}) can be obtained in the following form 
\begin{align}
&\chi_N[\bs{\xi},t]\notag\\
&=Z^{-1}  \int\mathcal{D}(\bar{c},c)\exp\left[i\int_{\mathcal{C}}dz_{1}dz_{2}\sum_{ij}\bar{c}_{i}(z_{1})G_{ij}^{-1}(z_{1},z_{2})c_{j}(z_{2})\right.\notag\\
&+\ci\int_{\mathcal{C}}dz\sum_{i\in A}\left[\bar{c}_{i}(z)\delta_\mathcal{C}\left(z,(t^+,+)\right)\xi_i-\bar{\xi}_i\delta_\mathcal{C}\left(z,(t,+)\right)c_{i}(z)\right]
\end{align}
where $G_{ij}(z_1,z_2)$ is the contour-ordered single-particle Green's function that encodes the details of the non-equilibrium process, e.g. for time-dependent Hamiltonian $H(t)=\sum_{ij}t_{ij}(t)c_i^\dagger c_j$
with time-dependent hopping $t_{ij}(t)$, $G_{ij}^{-1}(z_{1},z_{2})=\left((i\partial_{z_{1}}+\mu)\delta_{ij}-t_{ij}(t_1)\right)\delta_{\mathcal{C}}(z_{1}-z_{2})$ ($z_1=(t_1,\pm)$). We can integrate out the fermions to get
\begin{align}
\chi_N(\bs{\xi},t)&=\exp\left[\sum_{i,j\in A}\bar{\xi}_i\left\{-\ci G_{ij}\left((t,+),(t^+,+)\right)\right\}\xi_j\right]\nonumber \\
&=\exp\left[\sum_{ij\in A}\bar{\xi}_iC^T_{ij}(t)\xi_j\right]
\end{align}
The last line follows from the relation $G_{ij}\left((t,+),(t^+,+)\right)=G^\mathrm{T}(t,t^+)=\ci\langle c_j^\dagger(t)c_i(t)\rangle=\ci C^T_{ij}(t)$ involving the time-ordered Green's function $G^\mathrm{T}$. Using \eqn{eq:expS2-expression_t} and the characteristic function above, the second R\'{e}nyi entropy can be immediately evaluated by integrating out the auxiliary Grassmann variables $\bs{\xi}$ and $\bs{\eta}$ to get
\begin{align}\mylabel{eqn:S2-nI-t}
    S_A^{(2)}(t)	
	&=-\Tr\ln\left[(\I-C(t))^{2}+C(t)^{2}\right],
\end{align}
as discussed in \seccite{subsec:nI-formalism}.
The above is of the same form as obtained in \eqn{eqn:S2-nI-thermal} for the thermal case. This implies a similar recursion relation for higher R\'{e}nyi entropies as the one given in \eqn{eq:recurse_1} can be derived and solved in exactly the same way to give the final expression reported in \eqn{eq:Sn_nI} of the main text.


\section{Non-equilibrium field theory for R\'{e}nyi entropy in SYK model}\mylabel{app:keldysh-SYK}
The imaginary-time thermal field theoretic formulation discussed in \Seccite{sec:allmodels} for SYK model and its extensions allows us to access ground-state entanglement in these large-$N$ model. In this section we discuss the large-$N$ Schwinger-Keldysh formulation which will enable us to track the time evolution of entanglement entropy under non-equilibrium situations \cite{Eberlein2017,Sonner2017,Kourkoulou2017,Bhattacharya2018,Haldar2020,Zhang2019,Almheiri2019,Maldacena2019,Kuhlenkamp2020}. We demonstrate this by deriving the Swinger-Keldysh action and saddle-point equations for out-of-equilibrium evolution in the $SYK_q$ model (\Eqn{eq:SYKqham}) when the strength of interactions is varying with time, such that
\begin{align}
    \langle |J_{ijkl}|^2\rangle=J(t)^2/qN^{2q-1}(q!)^2.
\end{align}
Here $J(t)$ is an arbitrary function of time. Also, we take the initial density matrix as a thermal ensemble for the $SYK_q$ model prepared for some initial configuration of $J_{ijkl}$s. We use the path-integral derived in \Eqn{eqn:chi-Keldysh-thermal} and a modified form of the contour defined in \Eqn{eqn:CTC1-def} to develop the formulation. Instead, of stretching the real-time branches to $+\infty$ (see \Fig{fig:CTCs}(b)) we stop it at $t$, the time at which entanglement is to be measured.
Therefore, the new contour is now defined as 
\begin{align}\mylabel{eqn:CTC2-def}
\mathcal{C}=[t_0+\ci\beta,t_0)\cup [t_0,t]\cup(t,t_0],
\end{align}
where the imaginary-time contour remains same as that shown in \Fig{fig:CTCs}(b).

\rem{
Following this, we introduce disorder-replicas, perform disorder averaging like we did for the thermal-field theory and arrive at the following expression for $2-$nd R\'{e}nyi entropy
\begin{align}
    \exp[-\retwo(t)]=\frac{1}{Z^{2r}}\int\mathcal{D}(\bar{c}_{a},c_{a},\xi_{ia},\eta_{ia})e^{i\mathcal{S}[\bar{c}_{a},c_{a};\xi_{ia},\eta_{ia}]},
\end{align}
where
\begin{align}
    \mathcal{S}=&\int_{\mathcal{C}}dz_{1}\left[\sum_{i,\alpha a}\bar{c}_{i\alpha a}(z_{1})\ci\partial_{z_{1}}c_{i\alpha a}(z_{1})\right]\notag\\
    +&i\int_{\mathcal{C}}dz_{1,2}\sum_{\alpha\beta ab}\frac{J^{2}}{2qN^{2q-1}}\notag\\
    &\left(\sum_{ij}c_{i\alpha a}(z_{1})\bar{c}_{i\beta b}(z_{2})c_{j\beta b}(z_{2})\bar{c}_{j\alpha a}(z_{1})\right)^{q}\\&+\int_{\mathcal{C}}dz_{1}\sum_{i\in A,\alpha a}\left(\bar{c}_{i\alpha a}(z_{1})j_{i\alpha a}(z_{1})+\bar{j}_{i\alpha a}(z_{1})c_{i\alpha a}(z_{1})\right)
\end{align}
}
The steps are similar to the thermal case of \Seccite{sec:allmodels}, but with the imaginary-time $\tau$ generalized to the contour-variable $z$, and we end up with an intermediate replica-diagonal Keldysh-action
\begin{align}\mylabel{eq:EEK-action}
\mathcal{S}=&\left(-\sum_{i\in A,}\left[\begin{array}{cc}
\bar{\xi}_{i} & \bar{\eta}_{i}\end{array}\right]\left[\begin{array}{cc}
\frac{1}{2} & -\frac{1}{2}\\
\frac{1}{2} & \frac{1}{2}
\end{array}\right]\left[\begin{array}{c}
\xi_{i}\\
\eta_{i}
\end{array}\right]\right.\notag\\
&
+\int_{\mathcal{C}}dz\ \delta_\mathcal{C}(z_{-}-t)\left[\begin{array}{cc}
\bar{c}_{i1}(z) & \bar{c}_{i2}(z)\end{array}\right]\left[\begin{array}{c}
\xi_{i}\\
\eta_{i}
\end{array}\right]\notag\\
&-\left.\int_{\mathcal{C}}dz\ \delta_\mathcal{C}(z-t)\left[\begin{array}{cc}
\bar{\xi}_{i} & \bar{\eta}_{i}\end{array}\right]\left[\begin{array}{c}
c_{i1}(z)\\
c_{i2}(z)
\end{array}\right]\right)\notag\\
&+\int_{\mathcal{C}}dz_{1}\sum_{i,\sigma}\bar{c}_{i\sigma}(z_{1})
\ci\partial_{z_{1}}c_{i\sigma}(z_{1})\notag\\
&-\ci N\int dz_{1,2}\sum_{\sigma\sigma'}\left[\Sigma_{\sigma\sigma'}(z_{1},z_{2})
G_{\sigma'\sigma}(z_{2},z_{1})\right.
\notag\\
&\left.-\frac{J(z_{1})J(z_{2})}{2q}\left(G_{\sigma\sigma'}(z_{2},z_{1})\right)^{q}
\left(G_{\sigma'\sigma}(z_{1},z_{2})\right)^{q}\right],
\end{align}
where the contour delta function $\delta_\mathcal{C}(z_{-}-t)$ is nonzero when $z$ approaches the measurement time $t$ along the $-$ branch of the contour and $\delta_\mathcal{C}(z-t)$ is nonzero when $z$ approaches $t$ from the $+$ branch of the contour, see \Eqn{eqn:CTC2-def}. The large-$N$ field $G$ is upgraded to a contour version and is defined as
\begin{align}
    G_{\sigma'\sigma}(z_{2},z_{1})	=	\frac{\ci}{N}\sum_{i}\cb_{i\sigma}(z_{1})c_{i\sigma'}(z_{2}).
\end{align}
The rest of the symbols have the same meaning as the ones defined in \Eqn{eq:SYK-ZAr}. We emphasize, that the time dependence is explicitly encoded in the function $J(z)$, which is equal to $J(t)$ when $z\in[t_0,t]\cup(t,t_0]$ part of the contour, and a static quantity $J_0$ when $z\in[t_0+\ci\beta,t_0)$. If we integrate $\bs{\xi}$,$\bs{\eta}$ first, in \Eqn{eq:EEK-action}, and then the fermion-fields $\cb_{i\sigma}(z)$, $c_{i\sigma}(z)$, in that order, we arrive at the final expression for the entanglement-Keldysh action
\begin{align}\mylabel{eqn:EE-K-final-def}
    \mathcal{S}^\mathcal{C}&=
    -\ci \ln\det\left[-\ci(\ci\bs{\partial_{z}}-\bs{\Sigma})\right]\notag\\
    &-\ci p\ln\det\left[-\ci(\ci\bs{\partial_{z}}-\bs{\Sigma}+\ci \bs{M})\right]\notag\\
    &-\ci\int dz_{1,2}\sum_{\sigma\sigma'}\Big[ \Sigma_{\sigma\sigma'}(z_{1},z_{2})G_{\sigma\sigma'}(z_{2},z_{1})\notag\\
    &-\left. \frac{J(z_{1})J(z_{2})}{2q}G(z_{2},z_{1})^{q}G(z_{1},z_{2})^{q}\right]
\end{align}
in the large-$N$ limit. The symbols $\bs{\partial_{z}}$, $\bs{\Sigma}$ represents matrices having elements $\partial_{z_{1}}\delta_\mathcal{C}(z_{1}-z_{2})\delta_{\sigma\sigma'}$, $\Sigma_{\sigma\sigma'}(z_{1},z_{2})$ respectively, and the matrix $\bs{M}$ is defined as
\begin{align}\mylabel{eq:M-def_Keldysh}
    \bm{M}_{\sigma_{1}\sigma_{2}}(\tau_{1},\tau_{2})=\begin{bmatrix}\begin{array}{rr}
1 & 1\\
-1 & 1
\end{array}\end{bmatrix}\delta_\mathcal{C}(z_{1-}-t)\delta_\mathcal{C}(z_{2+}-t).
\end{align}
\Eqn{eqn:EE-K-final-def} is the time-dependent generalization of \Eqn{eq:action-before-freeE} that was derived in the context of thermal-field theory. And in the same manner, the saddle-point equations can be derived to yield
\begin{align}\mylabel{eq:SP-SYK-Keldysh}
\bs{G}	=&(1-p)\bs{\tilde{G}}+p\bs{g}\notag\\
\bs{\tilde{G}}	=&(\ci\bs{\partial_z}-\bs{\Sigma})^{-1}\notag\\
\bs{g}	=&(\ci\bs{\partial_z}-\bs{\bs{\Sigma}}+\ci\bs{M})^{-1}\notag\\
\Sigma_{\sigma\sigma'}(z_{1},z_{2})	=&J(z_1)J(z_2)G_{\sigma\sigma'}(z_{1},z_{2})^{q}G_{\sigma'\sigma}(z_{2},z_{1})^{q-1},    
\end{align}
the solutions for which, when plugged into \Eqn{eqn:EE-K-final-def} provides us with the expression for $2$-nd R\'{e}nyi entropy density at time $t$, i.e.
\begin{align}
    \retwo(t)=-\ci\mathcal{S}^\mathcal{C}(t)+2N^{-1}\ln{Z},
\end{align}
where $Z$ is the partition function describing the initial thermal-ensemble and can be computed from standard thermal-field theory for the SYK$_q$ model. The generalization of the formulation to other versions of the SYK model can be done in exactly the same manner and will be discussed elsewhere \cite{Haldar2020Renyi}. 
\section{Saddle point equations and entanglement free-energy for the models of FL and NFL-FL transition} \label{app:OtherModels}
In this appendix we provide the saddle-point equations and entanglement free energies for the rest of the large-$N$ modules introduced in \seccite{sec:allmodels}.

{\bf SYK model with quadratic hopping term:}
The saddle point equations for the interacting Fermi liquid defined in \eqn{eq:SYKFLham} are same as \eqn{eq:SP-SYK}, except the formula for the self-energy is now given by
\begin{align}\mylabel{eq:SP-SYKFL}
   \Sigma_{\sigma\sigma'}(\tau_{1},\tau_{2})	=&(-1)^{q+1}J^{2}G_{\sigma\sigma'}(\tau_{1},\tau_{2})^{q}
   G_{\sigma'\sigma}(\tau_{2},\tau_{1})^{q-1}\notag\\
   &+\thop^{2}G_{\sigma\sigma'}(\tau_{1},\tau_{2}).    
\end{align}
Also, the entanglement free energy becomes
\begin{align}
    &F_{EE}(p,\beta)\notag\\
    &=\frac{1}{2\beta}\left[p\ln\det\left(-\mathbf{g}\right)+(1-p)\ln\det\left(-\mathbf{\tilde{G}}\right)\right.\notag\\
	&-\int_{0}^{\beta}\dtau_{1,2}\sum_{\sigma=1,2}(-1)^{q}\frac{J^{2}}{2q}G_{\sigma'\sigma}(\tau_{2},\tau_{1})^{q}G_{\sigma\sigma'}(\tau_{1},\tau_{2})^{q}\notag\\
	&+\int_{0}^{\beta}\dtau_{1,2}\sum_{\sigma=1,2}\thop^{2}G_{\sigma'\sigma}(\tau_{2},\tau_{1})
	G_{\sigma\sigma'}(\tau_{1},\tau_{2})\notag\\
	&\left.-\int_{0}^{\beta}\dtau_{1,2}\sum_{\sigma=1,2}\Sigma_{\sigma\sigma'}(\tau_{1},\tau_{2})G_{\sigma'\sigma}(\tau_{2},\tau_{1})\right].
\end{align}

{\bf BA model for non Fermi liquid to Fermi liquid transition:}
We determine the saddle point equations for calculating R\'{e}nyi-entropy in the BA model (see \eqn{eq:BAModel}) for the following two subsystem choices -- (a) when the subsystem is chosen to be the SYK c-fermions and (b) when the subsystem is formed by the non-interacting $\psi$-fermions. In both cases $p$ represents the ratio of $\psi$-fermion sites ($N_\psi$) to $c$-fermion sites ($N_c$), i.e. $p=N_\psi/N_c$. The self-energies $\Sigma^{(c)}$, $\Sigma^{(\psi)}$, for both cases (a) and (b), are given by
\begin{align}
\Sigma^{(c)}_{\sigma\sigma'}(\tau_{1},\tau_{2})&=-J^{2}G^{(c)}_{\sigma\sigma'}(\tau_{1},\tau_{2})^2G^{(c)}_{\sigma'\sigma}(\tau_{2},\tau_{1})\notag\\
&+\sqrt{p}V^{2}G^{(\psi)}_{\sigma\sigma'}(\tau_{1},\tau_{2})\notag\\
\Sigma^{(\psi)}_{\sigma\sigma'}(\tau_{1},\tau_{2})&=\thop^{2}G^{(\psi)}_{\sigma\sigma'}(\tau_{1},\tau_{2})
+\frac{V^{2}}{\sqrt{p}}G^{(c)}_{\sigma\sigma'}(\tau_{1},\tau_{2}).
\end{align}
The equations for obtaining the Green's functions from the self-energies are 
\begin{align}
\bs{G}^{(c)}&=-\left(\bs{\del}_{\tau}+\bs{\Sigma}^{(c)}+M\right)^{-1}\notag\\
\bs{G}^{(\psi)}&=-\left(\bs{\del}_{\tau}+\bs{\Sigma}^{(\psi)}\right)^{-1}
\end{align}
for case (a), and
\begin{align}
\bs{G}^{(c)}&=-\left(\bs{\del}_{\tau}+\bs{\Sigma}^{(c)}\right)^{-1}\notag\\
\bs{G}_{(\psi)}&=-\left(\bs{\del}_{\tau}+\bs{\Sigma}^{(\psi)}+M\right)^{-1}
\end{align}
for case (b), respectively. The entanglement free-energy is again same for both the cases and is given by
\begin{align}
&F_{EE}(p,\beta)\notag\\
=&\frac{1}{2\beta(1+p)}{\Big[}\ln\det\left(-\bs{G}^{(c)}\right)+p\ln\det\left(-\bs{G}^{(\psi)}\right)\notag\\
&-\int_{0}^{\beta}\dtau_{1,2}\sum_{\sigma\sigma'}\frac{J^{2}}{4}G_{\sigma'\sigma}^{(c)}(\tau_{2},\tau_{1})^{2}G_{\sigma\sigma'}^{(c)}(\tau_{1},\tau_{2})^{2}\notag\\
&-\int_{0}^{\beta}\dtau_{1,2}\sum_{\sigma\sigma'}\Sigma_{\sigma\sigma'}^{(c)}(\tau_{1},\tau_{2})G_{\sigma'\sigma}^{(c)}(\tau_{2},\tau_{1})\notag\\
&+p\int_{0}^{\beta}\dtau_{1,2}\sum_{\sigma\sigma'}\frac{\thop^{2}}{2}G_{\sigma'\sigma}^{(\psi)}(\tau_{2},\tau_{1})G_{\sigma\sigma'}^{(\psi)}(\tau_{1},\tau_{2})\notag\\
&-p\int_{0}^{\beta}\dtau_{1,2}\sum_{\sigma\sigma'}\Sigma_{\sigma\sigma'}^{(\psi)}(\tau_{1},\tau_{2})G_{\sigma'\sigma}^{(\psi)}(\tau_{2},\tau_{1})\notag\\
&\left.+\sqrt{p}\int_{0}^{\beta}\dtau_{1,2}\sum_{\sigma\sigma'}V^{2}G_{\sigma'\sigma}^{(c)}(\tau_{2},\tau_{1})G_{\sigma\sigma'}^{(\psi)}(\tau_{1},\tau_{2})\right].
\end{align}
\section{Coupling constant integration method to compute grand potentials}\mylabel{app:CouplingConstInt}
For the $n$-th R\'{e}nyi entropy we need to obtain the $n$-th grand potential, which is defined as
\begin{align*}
\Omega^{(n)}(\lambda)= & -T\ln Z^{(2)}(\lambda)=-T\ln\mathrm{Tr}_{A}Z_{A}^{n}(\lambda)\\
Z_{A}(\lambda)= &\mathrm{Tr}_{B}e^{-\beta\left(H_{0}+\lambda H_{1}\right)}
\end{align*}
Here, $H_{0}$ is a reference Hamiltonian, typically the
non-interacting part, whose
grand potential can be obtained easily. $H_{1}$ is the
interacting part and we have multiplied by a real variable $\lambda$,
which will be integrated over; The Hamiltonian $H_{\lambda}=H_{0}+\lambda H_{1}$and
$\lambda=0$ is the non-interacting limit and $\lambda=1$ is the
interacting Hamiltonian of interest. $n=2$ correspond to
the second R\'{e}nyi potential and $n=1$ the usual grand potential. We
obtain
\begin{align}
\partial_\lambda \Omega^{(n)}(\lambda)= & \frac{-nT}{Z^{(n)}(\lambda)}\mathrm{Tr}_{A}\left[Z_{A}^{n-1}(\lambda)\left(\mathrm{Tr}_{B}\partial_\lambda e^{-\beta(H_{0}+\lambda H_{1})}\right)\right],
\end{align}
where $\partial_\lambda=\partial/\partial\lambda$. It can be easily shown that \cite{FetterBook}, $\partial_\lambda\exp[-\beta(H_{0}+\lambda H_{1})]=  -\beta \exp[-\beta(H_{0}+\lambda H_{1})]H_{1}$, and hence
\begin{align}
\partial_\lambda\Omega^{(n)}(\lambda)= & \frac{n}{\lambda Z^{(n)}(\lambda)}\mathrm{Tr}_{A}\left[Z_{A}^{n-1}(\lambda)\right.\nonumber\\
&\left.\left(\mathrm{Tr}_{B}e^{-\beta(H_{0}+\lambda H_{1})}\lambda H_{1}\right)\right].
\end{align}
For $n=2$, using the trace formula (\Eqn{eq:Tr-FG-exp-DN}) and integrating both sides over $\lambda$ from 0 to 1 we get
\begin{align}
\Omega^{(2)}(\lambda)= &\Omega^{(2)}(0)\nonumber \\
&+\int_0^1\frac{d\lambda}{\lambda}\frac{2}{Z^{(2)}(\lambda)}\int d^{2}(\bs{\xi},\bs{\eta})\left\{f_N(\bs{\xi},\bs{\eta})\right.\nonumber \\
&\left.\mathrm{Tr}\left[e^{-\beta H_\lambda}D_N(\bs{\xi})\right]\mathrm{Tr}\left[e^{-\beta H_\lambda}\lambda H_{1}D_N(\bs{\eta})\right]\right\},
\end{align}
where $\Omega^{(2)}(0)$ is the R\'{e}nyi grand potential for the non-interacting
system. As in \Eqn{eq:Z2-1}, one can construct a path
integral representation for the above using the generating function
$Z^{(2)}(\lambda)$ and obtain
\begin{align}
\Omega^{(2)}
= & \Omega^{(2)}(0)+\int_{0}^{1}\frac{d\lambda}{\lambda}\left\langle \lambda U\sum_{i}n_{i\uparrow\alpha}(0)n_{i\downarrow\alpha}(0)\right\rangle _{Z^{(2)}(\lambda)}. \label{eq:GrandPotRenyi_1}
\end{align}
Here $\alpha$ is the entanglement replica index. The important point to note
here is that the expectation of $\lambda H_{1}$ with respect
to $Z^{(2)}$ has to be calculated at the same time where $D_N(\bs{\xi})$
is inserted in the path integral i.e. at $\tau=0$ in our case. Naively, it seems that the above
requires the evaluation of a four-point function. However one can
write $\left\langle \lambda U\sum_{i}n_{i\uparrow\alpha}(0)n_{i\downarrow\alpha}(0)\right\rangle $
in terms of the single-particle Green's function. To show this, we use the Heisenberg equation of motion \cite{FetterBook}
\begin{align}
&\frac{dc_{i\sigma}(\tau)}{d\tau}= [H,c_{i\sigma}(\tau)]= -\sum_{j}t_{ij}c_{j\sigma}(\tau)+\mu c_{i\sigma}(\tau)\nonumber\\
&-\lambda U\left(n_{i\uparrow}(\tau)c_{i\downarrow}(\tau)\delta_{\sigma\downarrow}+n_{i\downarrow}(\tau)c_{i\uparrow}(\tau)\delta_{\sigma\uparrow}\right)
\end{align}
From the above we can easily show that
\begin{align}
&\left\langle \lambda U\sum_{i}n_{i\uparrow\alpha}(0)n_{i\downarrow\alpha}(0)\right\rangle _{Z^{(2)}(\lambda)}\nonumber\\
&=\frac{1}{2}\lim_{\tau'\to\tau^{+}}\sum_{ij\sigma}\left[(\partial_{\tau}-\mu)\delta_{ij}+t_{ij}\right]G_{j\sigma,i\sigma}^{(\lambda)}(\tau,\tau').
\end{align}
Using the above in \Eqn{eq:GrandPotRenyi_1}, we obtain the expression for second R\'{e}nyi grand potential in \Eqn{eq:GrandPotRenyi} in \Seccite{subsec:DMFT}. Also, the usual thermodynamic grand potential ($n=1$) is given by
\begin{align}
\Omega= & \Omega(0)+\int_{0}^{1}\frac{d\lambda}{\lambda}\left\langle \lambda U\sum_{i}n_{i\uparrow}(\tau)n_{i\downarrow}(\tau)\right\rangle _{Z(\lambda)},
\end{align}
and one can obtain similar expression in terms of the Green's function as shown in \Eqn{eq:GrandPotThermo} in \Seccite{subsec:DMFT}.

\section{Proof of maximal entanglement in SYK model for \emph{small} subsystems}\mylabel{sec:vol-laws-proof}
In this section, we analytically prove that the SYK model is maximally entangled with second R\'{e}nyi entropy $S_A^{(2)}=p\ln(2)$ for $p\to 0$, i.e. when the subsystem size becomes vanishing fraction of the total system size. To this end, we start from \eqn{eq:EE-action}, and instead of integrating out the $\bs{\xi}$, $\bs{\eta}$ variables first, we integrate the $c$-fermions and then the $\bs{\xi}$, $\bs{\eta}$ variables. This leads to $\overline{\Tr[Z_A^2]^r}=\int \mathcal{D}(\cb,c,\Sigma,G)e^{-Nr\mathcal{S}[\Sigma,G]}$ and an effective action under (disorder) replica symmetric and diagonal ansatz 
\begin{align}
\mathcal{S}&=-\ln\mathrm{det}(\partial_\tau+\Sigma)+S_G/r-p\ln\mathcal{K}\nonumber\\
\mathcal{K}&=2\left[(C_{11}-1/2)(C_{22}-1/2)-(C_{12}+1/2)(C_{21}-1/2)\right].
\end{align}
 Here $r$ is the number of disorder replicas, $S_G$ is obtained from \eqn{eq:action-before-freeE} by plugging in the ansatz $G,\Sigma\propto \delta_{ab}$ and $C_{\alpha\gamma}=\tilde{G}_{\alpha\gamma}(0,0^+)$ (see below for definition).
The above action produces an alternate (but mathematically equivalent to \eqn{eq:SP-SYK}) set of saddle-point equations 
\begin{align}\mylabel{eq:SP-SYK-1}
\bs{\tilde{G}}	=&-(\bs{\deltau}+\bs{\Sigma})^{-1}\notag\\
\bs{G}=&\bs{\tilde{G}}-p\frac{\delta \ln\mathcal{K}}{\delta \Sigma} \nonumber\\
\Sigma_{\sigma\sigma'}(\tau_{1},\tau_{2})	=&(-1)^{q+1}J^{2}G_{\sigma\sigma'}(\tau_{1},\tau_{2})^{q}G_{\sigma'\sigma}(\tau_{2},\tau_{1})^{q-1},    
\end{align}
When $p\to0$, the
the above saddle-point Green's functions are same as that of the original
 SYK saddle point in a thermal ensemble and $\tilde{G}_{\alpha\gamma}=G_{\alpha\gamma}=G\delta_{\alpha\gamma}$. Also $C_{\alpha\gamma}=n\delta_{\alpha\gamma}$ with $n=\langle c_i^\dagger (0)c_i(0)\rangle$, the fermion density per site. Hence using \eqn{eq:S2pbetadef} the R\'{e}nyi entropy density of the subsystem $A$ is simply given by 
\begin{align}
S^{(2)}(p)= & -p\ln\left[(1-n)^{2}+n^{2}\right].
\end{align}
For half filling, we get $S^{(2)}=p\ln(2)$, i.e. the maximum entanglement possible for spin-less fermions.

\section{Numerical techniques}\mylabel{sec:numerics}
In this section, we provide the details of the numerical techniques used to arrive at the results of \seccite{sec:Results}. The section contains two parts: the first part gives the details for finite-$N$ calculation and the second discusses the numerical solution of the saddle-point equations that appear in the main text.
\subsection{Finite-N numerics}\mylabel{sec:finite-N-num}
We write the thermal-density matrix, $\rho=\exp(-\beta H)/Z$, in the eigen-basis ($\{|\psi_\alpha\rangle\}$) of the Hamiltonian $H$ as follows
\begin{align}
	\rho = \sum_{\alpha} p_{\alpha} |\psi_{\alpha}\rangle \langle \psi_{\alpha}|,
\end{align}
where the probability $p_\alpha$ are calculated using
\begin{align}
p_\alpha=\exp(-\beta E_\alpha)/Z.
\end{align}
The many-body energies, $E_\alpha$, and the eigenstates, $|\psi_\alpha\rangle$, are obtained by exact diagonalization (ED) of the Hamiltonian $H$. The partition function, $Z$, for a given temperature $T$($=\beta^{-1}$) is calculated from the energies $E_\alpha$ using
\begin{align}
Z=\sum_\alpha \exp(-\beta E_\alpha).
\end{align}
We perform our calculations using the grand-canonical ensemble in order to include contribution from all number sectors in the Fock-space for fermions. The dimension of the Hilbert-space is $2^{N}$ with $N$ being the total number of fermion flavors/sites in the theory. The reduced density matrix ($\rho_A$) for the subsystem $A$, having $N_A$ sites, can be calculated from the reduced density matrices of the individual states $|\psi_\alpha\rangle$, as follows
\begin{align}\mylabel{eqn:red-rhoA-eigen}
\rho_A=\Tr_B\rho=\sum_\alpha p_\alpha\Tr_B\left[|\psi_{\alpha}\rangle \langle \psi_{\alpha}|\right],
\end{align}
where $\Tr_B$ represent the trace over the rest of the system, $B$, having $N-N_A$ sites. We define the  reduced density matrices of the individual state $|\psi_\alpha\rangle$ as 
\begin{align}\mylabel{eqn:red-rho-eigen}
\rho^{(\alpha)}_A=\Tr_B\left[|\psi_{\alpha}\rangle \langle \psi_{\alpha}|\right],
\end{align}
which can be calculated by rewriting $|\psi_\alpha\rangle$ in the form
\begin{align}
   |\psi_{\alpha}\rangle = \sum_{{i \in A},{j \in B}} \psi^{(\alpha)}_{i,j} |i\rangle_A \otimes |j\rangle_B\equiv\sum_{{i \in A},{j \in B}} \psi^{(\alpha)}_{i,j} |i_A;j_B\rangle, 
\end{align}
where the vectors $|i\rangle_A$ ($i=1,\cdots,2^{N_A}$) span the Hilbert space for the subsystem $A$, while $|j\rangle_B$ ($j=1,\cdots,2^{N-N_A}$) span that of $B$, the rest of the system. The amplitude $\psi^{(\alpha)}_{i,j}$ can be viewed as a matrix with dimension $2^{N_A}\times 2^{N-N_A}$. The tracing over the $B$ subsystem (see \eqn{eqn:red-rho-eigen}) can be performed to give
\begin{align}
	(\rho^{(\alpha)}_{A})_{i,i'} =&\sum_{j\in B} \langle i_A; j_B| \psi_{\alpha}\rangle \langle \psi_{\alpha}|i'_A; j_B\rangle\notag\\
	&= \sum_{j\in B} \psi^{(\alpha)}_{i,j}(\psi^{(\alpha)}_{i',j})^*,
\end{align} 
where $(\rho^{(\alpha)}_{A})_{i,i'}$ is the $i,i'$-th element of the reduced density matrix $\rho^{(\alpha)}_A$. The reduced density matrix for the subsystem $A$ can now be calculated using \eqn{eqn:red-rhoA-eigen}, i.e. 
\begin{align}
\rho_A=\sum_\alpha p_\alpha \rho^{(\alpha)}_A,
\end{align}
and $2$-nd R\'{e}nyi entropy, $\retwo$, can finally be evaluated using
\begin{align}
\retwo=-\overline{\ln\left(\Tr_A[\rho^2_A]\right)}.
\end{align}
The $\overline{\cdots}$, appearing above, denotes averaging over disorder realizations of $J_{ijkl}$s (see \eqn{eq:SYKqham}). We performed averaging over $2000$ disorder realizations for calculations involving $N=10$ fermionic sites and averaging over $250$ realizations for system involving $N=12$ sites. The reduced density matrix $\rho_A$ was calculated for each new disorder realization.

\subsection{Solution of saddle-point equations}\mylabel{sec:sp-num}
We demonstrate our numerical approach by using the SYK$_q$ model, defined in \eqn{eq:SYKqham}, as a prototype. The saddle-point equations for which are given in \eqn{eq:SP-SYK}. We reiterate these equation here for ease of access
\begin{align}
\bs{G}	=&(1-p)\bs{\tilde{G}}+p\bs{g}\mylabel{eq:SP-l1}\\
\bs{\tilde{G}}	=&-(\bs{\deltau}+\bs{\Sigma})^{-1}\mylabel{eq:SP-l2}\\
\bs{g}	=&-(\bs{\deltau}+\bs{\bs{\Sigma}}+\bs{M})^{-1}\mylabel{eq:SP-l3}\\
\Sigma_{\sigma\sigma'}(\tau_{1},\tau_{2})	=&(-1)^{q+1}J^{2}G_{\sigma\sigma'}(\tau_{1},\tau_{2})^{q}G_{\sigma'\sigma}(\tau_{2},\tau_{1})^{q-1}\mylabel{eq:SP-l4}.    
\end{align}
We discretize the domain $[0,\beta)$ for imaginary-time $\tau$ into $N_\tau$ segments. Since, the matrix $\bs{M}$, appearing above, breaks time-translation symmetry (see \eqn{eq:M-def}), we represent the Green's functions ($\bs{G}$, $\bs{g}$ etc.) and self-energy ($\bs{\Sigma}$) as matrices having dimensions $2N_\tau\times 2N_\tau$. The factor $2$ accounts for the two copies of the reduced density matrix in $2$-nd R\'{e}nyi-entropy (see \eqn{eq:S2-def-thermal-SYK}), evaluating $n$-th R\'{e}nyi-entropy will require $n N_\tau\times n N_\tau$ matrices.
The time derivative is also represented as a matrix using the finite-difference relations given in \eqn{eqn:finite-diff-def}. The anti-periodic boundary conditions for the Green's function, i.e. $G(\tau_1+\beta,\tau_2)=-G(\tau_1,\tau_2)=G(\tau_1,\tau_2+\beta)$ etc., are incorporated into the matrix representation of the derivative operator as well. With this setup, the saddle-point equations can be solved by starting from an initial guess for the Green's function $\bs{\tilde{G}}$, $\bs{g}$. Using which, the self-energy $\bs{\Sigma}$ can be determined using \eqn{eq:SP-l4}. A new ``corrected" $\bs{\tilde{G}}$, $\bs{g}$ can then found by performing the inverses in \eqn{eq:SP-l2} and \eqn{eq:SP-l3} numerically. The process can then be iterated till sufficient numerical convergence has been achieved.
 
\bibliography{SYK}

\ifdefined\makeSM




%
%
\fi
\end{document}